\documentclass[11pt,a4paper]{article}

\usepackage[utf8]{inputenc}
\usepackage[T1]{fontenc}
\usepackage[margin=2.5cm]{geometry}
\usepackage{amsmath,amssymb,amsthm,mathtools,bm}
\usepackage{graphicx}
\usepackage{booktabs,array,tabularx,longtable,multirow}
\usepackage{enumitem}
\usepackage[table]{xcolor}
\usepackage[colorlinks=true,linkcolor=blue!60!black,citecolor=blue!60!black,urlcolor=blue!60!black]{hyperref}
\usepackage{natbib}
\usepackage{setspace}
\usepackage{fancyhdr}
\usepackage{titlesec}
\usepackage{float}
\usepackage{algorithm}
\usepackage{algpseudocode}
\usepackage{caption}
\usepackage{subcaption}
\usepackage{tikz}
\usetikzlibrary{arrows.meta,positioning}

\definecolor{accent}{RGB}{19,61,122}
\titleformat{\section}{\Large\bfseries\color{accent}}{\thesection}{1em}{}
\titleformat{\subsection}{\large\bfseries\color{accent!80}}{\thesubsection}{1em}{}
\titleformat{\subsubsection}{\normalsize\bfseries\color{accent!60}}{\thesubsubsection}{1em}{}

\newcommand{\paperdate}{April 2026}

\newcommand{\R}{\mathbb{R}}

\newcommand{\diag}{\mathrm{diag}}
\newcommand{\balpha}{\bm{\alpha}}
\newcommand{\clip}{\mathrm{clip}_{\mathrm{SR}}}

\begin{document}
\thispagestyle{empty}

\begin{center}
{\LARGE\bfseries\color{accent}
Global Persistence, Local Residual Structure:\\[6pt]
Forecasting Heterogeneous Investment Panels\par}
\vspace{1cm}
{\large
Oleg Roshka\\[6pt]
}
{\normalsize
Independent Researcher\\[6pt]
\paperdate
}
\end{center}

\vspace{0.5cm}

\begin{abstract}
\noindent
On a 93-actor quarterly panel mixing macro indicators, institutional data, and firm-level investment ratios, global factor augmentation degrades prediction for actor subgroups whose dynamics are misrepresented by the shared basis. A two-stage architecture---global pooled AR(1) for shared persistence, block-specific local models for residual dynamics---improves full-panel out-of-sample $R^2$ from 0.630 to 0.677 ($\Delta = +0.047$, CI $[+0.036, +0.058]$, 10/10 windows, placebo $p \leq 0.001$). A held-out decade test (block partition frozen on 2005--2014 data, evaluated on unseen 2015--2024 windows) confirms the gain ($\Delta = +0.050$, 10/10), and a stratified placebo that fixes the macro/firm data-type split and permutes only firm-sector assignments corroborates ($z = 7.25$, $p \leq 0.001$). Cross-regime replication on a 109-actor UK/EU heterogeneous panel ($\Delta = +0.017$, 8/8 windows) and a combined US + UK/EU panel of 202 actors ($\Delta = +0.030$, placebo $z = 9.68$ --- exceeding the original US-only $z = 7.82$) confirms the architecture transfers across regimes. A 146-firm CapEx/Assets robustness check refines the scope condition: the gain depends on cross-sectional dispersion in autoregressive structure, which data-type heterogeneity reliably produces but which is also present in firm-only panels under suitable ratio choices.
\end{abstract}

\vspace{0.3cm}
\noindent\textbf{Keywords:} heterogeneous panels, factor augmentation, cross-sectional forecasting, investment dynamics, block-specific estimation, panel heterogeneity

\noindent\textbf{JEL Classification:} C32, C38, C53, E22, G11

\newpage

\section{Introduction}
\label{sec:intro}

Factor models are ubiquitous in cross-sectional financial forecasting. The standard practice is to estimate a global basis from the training panel, project all actors onto it, and model dynamics in the shared factor space. Principal component analysis \citep{stock2002forecasting,bai2002determining}, dynamic factor models \citep{doz2012quasi}, and spectral decompositions \citep{schmid2010dynamic} all follow this template. The implicit assumption is that a single low-rank basis serves all actors in the panel equally well.

On a multilayer investment panel containing 93 actors---macro indicators, institutional data, and firm-level ratios---global spectral augmentation improves average predictive $R^2$ by 3.6~percentage points over per-actor AR(1) (Table~\ref{tab:augmentation}), yet simultaneously \emph{reduces} $R^2$ for the diversified-sector subgroup by 2.3~pp below the pooled-only baseline that uses no augmentation at all. The global basis captures cross-sector rotational dynamics that are noise for within-sector prediction; applying it to heterogeneous actors injects interference rather than extracting signal.

The diagnosis leads to a direct architectural fix. Shared persistence---the dominant source of quarterly cross-sectional predictability---benefits from global pooling because it is common across actor types. But the residual cross-sectional dynamics, which the second stage of a two-stage pipeline is designed to capture, are \emph{block-specific}: technology and healthcare actors co-move in patterns that are disrupted when mixed with macro indices and diversified-sector firms. Estimating residual dynamics locally, within economically motivated blocks, avoids the cross-block interference.

The resulting heterogeneity-aware architecture---global Stage~1 (pooled AR(1) with fixed effects) plus block-specific local Stage~2 (PCA+ridge on within-block residuals)---improves full-panel out-of-sample $R^2$ from 0.630 to 0.677 ($\Delta = +0.047$, CI $[+0.036, +0.058]$, 10/10 rolling windows positive). A placebo benchmark using 1{,}000 random block partitions of identical sizes produces a mean gain of $-0.004$; the real partition's gain exceeds the placebo distribution by 7.82 standard deviations ($p < 0.001$). The block structure captures genuine panel heterogeneity, not statistical noise.

The gain does not stem from a better estimation method within the linear class. Across 9~linear specifications spanning three complexity classes, Dynamic Mode Decomposition, PCA, and Ridge regression produce statistically indistinguishable predictions at matched effective complexity (pairwise forecast-error correlations $\rho$ between $0.969$ and $0.990$). Per-actor gradient boosting breaks the linear ceiling ($R^2 = 0.661$) but falls short of the mixture ($R^2 = 0.677$); applying the same block decomposition to GBM ($R^2 = 0.657$) does not close the gap ($\Delta = -0.020$, 1/10 windows), showing that the mixture advantage combines block-specific estimation with low-rank factor extraction that per-actor non-linear models cannot replicate.

Several alternative explanations for the ceiling are tested and ruled out. The gain is not from forecasting the basis rotation (the global spectral basis rotates at $49^\circ$/quarter at $K{=}8$, but this rotation is temporally unpredictable: ACF(1) $= -0.07$, Ljung-Box $p = 0.59$). It is not from reformulating the prediction target (the ceiling is invariant across seven target transformations). It is not from conditional gating (augmentation is unconditionally beneficial---five gating policies all perform worse than always-on). Block-specific decomposition is the only surviving path to improvement.

The cross-panel validation sharpens the scope of the finding. The mixture architecture is null on a 146-firm CapEx/Revenue panel ($\Delta = -0.003$) and a 270-actor multi-ratio panel ($\Delta = +0.001$), while improving the 93-actor multilayer panel by $+0.047$. A robustness check on the same 146 firms using CapEx/Assets---the only methodological change---yields $\Delta = +0.013$ at a 3-year training window and $\Delta = +0.018$ at a 5-year window (both $p < 0.001$, 10/10 windows positive). The operative scope condition is therefore not strict data-type heterogeneity but rather sufficient cross-sectional dispersion in the autoregressive structure of the actors in the panel; data-type heterogeneity reliably produces this dispersion, but firm-only panels with appropriate ratio choices can also satisfy it.

Section~\ref{sec:cross_regime} reports a cross-regime extension. A 109-actor UK/EU heterogeneous panel (sampled 2011Q2--2025Q4 from EODHD) yields $\Delta = +0.017$ at a 3-year training window (NW-HAC bw=1 one-sided $p < 0.001$, 8/8 windows positive, placebo $z = 2.31$). A combined US + UK/EU panel of 202 actors over the overlap window yields $\Delta = +0.030$ (NW-HAC bw=1 one-sided $p < 0.0001$, 8/8 windows positive); the combined-panel placebo $z = 9.68$ exceeds the original US-only $z = 7.82$---the architecture replicates more cleanly across regimes than within the original US sample.

Beyond forecasting, the analysis yields a geometric description of panel heterogeneity. At a common rank $K{=}4$, the global spectral basis rotates at $31.5^\circ$/quarter (geodesic distance on the Grassmannian); within sector blocks, rotation is lower ($14$--$28^\circ$). A matched-size random sub-panel control indicates this reduction is partly a sample-size effect (empirical $p > 0.05$ for all blocks), so the geodesic evidence is treated as descriptive context, not as a demonstrated mechanism.

The paper proceeds as follows. Section~\ref{sec:data} describes the data and evaluation protocol. Section~\ref{sec:architecture} presents the two-stage architecture, method equivalence, and the heterogeneity-aware extension. Section~\ref{sec:results} presents the main empirical results. Section~\ref{sec:validation} provides placebo, cross-panel, and economic validation. Section~\ref{sec:cross_regime} reports the UK/EU extension and combined US+UK/EU panel results. Section~\ref{sec:falsification} summarises what does not work and why. Section~\ref{sec:discussion} discusses implications and limitations.

\section{Data and Setting}
\label{sec:data}

\subsection{The 93-Actor Multilayer Panel}

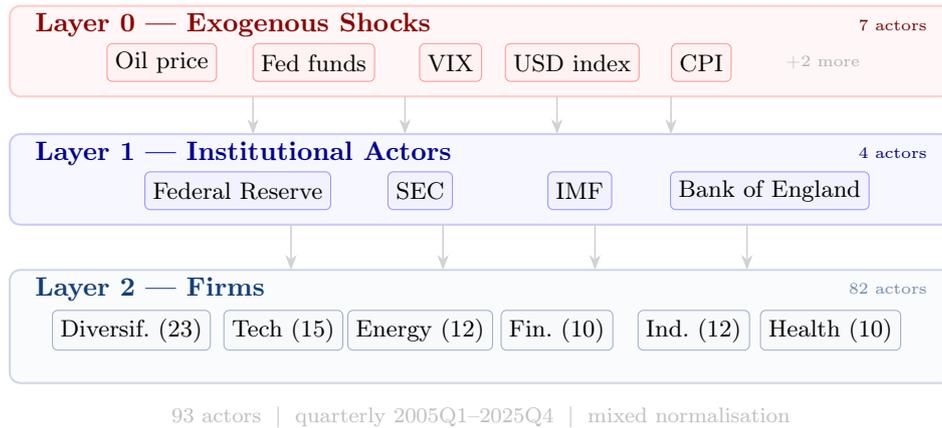
\begin{figure}[ht]
\centering
\begin{tikzpicture}[
  a0/.style={rectangle, draw=red!40, fill=red!5, rounded corners=2pt,
             font=\footnotesize, minimum height=0.5cm, inner sep=3pt},
  a1/.style={rectangle, draw=blue!40, fill=blue!5, rounded corners=2pt,
             font=\footnotesize, minimum height=0.5cm, inner sep=3pt},
  a2/.style={rectangle, draw=accent!40, fill=accent!3, rounded corners=2pt,
             font=\footnotesize, minimum height=0.5cm, inner sep=3pt},
  arr/.style={-{Stealth[length=2mm]}, color=gray!35, semithick},
  band/.style={rounded corners=4pt, thick}
]
\fill[red!3, band, draw=red!20] (-6.2, 3.05) rectangle (6.2, 4.25);
\node[font=\small\bfseries, text=red!55!black, anchor=west] at (-6.0, 4.0)
  {Layer 0 --- Exogenous Shocks};
\node[font=\tiny, text=red!45!black, anchor=east] at (6.0, 4.0) {7 actors};
\node[a0] at (-4.2, 3.5) {Oil price};
\node[a0] at (-2.2, 3.5) {Fed funds};
\node[a0] at (-0.4, 3.5) {VIX};
\node[a0] at (1.2, 3.5)  {USD index};
\node[a0] at (2.9, 3.5)  {CPI};
\node[font=\tiny, text=gray!60] at (4.5, 3.5) {+2 more};
\fill[blue!3, band, draw=blue!20] (-6.2, 1.35) rectangle (6.2, 2.55);
\node[font=\small\bfseries, text=blue!55!black, anchor=west] at (-6.0, 2.3)
  {Layer 1 --- Institutional Actors};
\node[font=\tiny, text=blue!45!black, anchor=east] at (6.0, 2.3) {4 actors};
\node[a1] at (-3.2, 1.8) {Federal Reserve};
\node[a1] at (-0.8, 1.8) {SEC};
\node[a1] at (1.3, 1.8)  {IMF};
\node[a1] at (3.8, 1.8)  {Bank of England};
\fill[accent!2, band, draw=accent!20] (-6.2, -0.75) rectangle (6.2, 0.75);
\node[font=\small\bfseries, text=accent, anchor=west] at (-6.0, 0.5)
  {Layer 2 --- Firms};
\node[font=\tiny, text=accent!60, anchor=east] at (6.0, 0.5) {82 actors};
\node[a2] at (-4.6, -0.05) {Diversif.\ (23)};
\node[a2] at (-2.6, -0.05) {Tech (15)};
\node[a2] at (-0.8, -0.05) {Energy (12)};
\node[a2] at (1.0, -0.05)  {Fin.\ (10)};
\node[a2] at (2.8, -0.05)  {Ind.\ (12)};
\node[a2] at (4.6, -0.05)  {Health (10)};
\foreach \x in {-3.0, -1.0, 1.0, 2.5}
  \draw[arr] (\x, 3.05) -- (\x, 2.55);
\foreach \x in {-2.5, -0.5, 1.5, 3.5}
  \draw[arr] (\x, 1.35) -- (\x, 0.75);
\node[font=\scriptsize, text=gray!50, align=center] at (0, -1.2)
  {93 actors \;\textbar\; quarterly 2005Q1--2025Q4 \;\textbar\; mixed normalisation};
\end{tikzpicture}
\caption{Multilayer actor hierarchy. Layer~0 comprises exogenous macro shocks (FRED min-max normalised); Layer~1 comprises institutional intermediaries; Layer~2 comprises US-listed firms grouped by sector (cross-sectional percentile ranks). The heterogeneous composition and mixed normalisation create the cross-block interference that motivates the mixture architecture (Section~\ref{sec:results}).}
\label{fig:hierarchy}
\end{figure}

The primary panel contains 93 actors observed quarterly from 2005Q1 to 2025Q4 (84~quarters). Actors span three layers (Figure~\ref{fig:hierarchy}) that reflect the hierarchical structure of the investment system:

\begin{itemize}[nosep]
  \item \textbf{Layer~0 (7 macro shocks):} Global commodity prices (Brent, WTI), policy rates (Federal Funds, USD index), and volatility indicators (VIX, credit spreads). Normalised via FRED min-max scaling using full-sample bounds; a robustness check with recursive expanding-window bounds confirms this does not affect the results (Appendix~\ref{app:fred}). These series are trending with high persistence ($\rho \approx 0.88$).
  \item \textbf{Layer~1 (4 institutional actors):} Central banks, regulators, and international organisations. Mixed normalisation: each institutional series uses its own native scaling (policy-rate levels, index values), with no cross-sectional ranking applied. Like the Layer~0 macro series, these are available in real time and involve no look-ahead in their normalisation.
  \item \textbf{Layer~2 (82 firms):} US-listed firms across six sectors (energy, diversified, technology, financials, industrials, healthcare), selected as the intersection of S\&P~500 constituents with complete quarterly data over the sample period (a balanced-panel requirement that induces survivorship bias toward persistent, large-cap firms; see Limitation~7). Intensity values are contemporaneous within-quarter cross-sectional percentile ranks of investment ratios (CapEx/Assets, Revenue growth)---each quarter's ranks use only that quarter's cross-section, so no future information enters the ranking. However, because the ranking pool consists exclusively of the 82 balanced-panel survivors, the cross-sectional distribution at every quarter is conditioned on future survival (see Limitation~7). These series are mean-reverting ($\rho \approx 0.60$).
\end{itemize}

The mixed normalisation---trending min-max series for macro actors and mean-reverting cross-sectional ranks for firms---reflects fundamentally different data-generating processes. This difference in temporal dynamics across layers is a root cause of the heterogeneity effects documented in Section~\ref{sec:results}. We state it here so the reader can anticipate the finding.

Seven sector labels provide the basis for block structure: diversified~(23), technology~(15), energy~(12 firms + 2 Layer~0 macro shocks), financials~(10 firms + 2 Layer~1 institutions), industrials~(12), healthcare~(10), and macro~(7). Table~\ref{tab:panel_stats} reports firm-layer counts only (82 total); the full panel of 93 includes the Layer~0/1 actors whose sector labels span layers. The disjoint mixture partition assigns all Layer~0/1 actors to a separate macro/institutional block regardless of their sector label. Table~\ref{tab:panel_stats} summarises the panel structure.

\begin{table}[ht]
\centering
\caption{Panel descriptive statistics. Persistence $\rho$ is the mean lag-1 autocorrelation within each layer.}
\label{tab:panel_stats}
\small
\begin{tabular}{lccccl}
\toprule
Layer / Sector & $N$ & Quarters & Normalisation & Persistence $\rho$ & Data type \\
\midrule
Layer 0 (Macro) & 7 & 84 & FRED min-max & 0.88 & Trending indices \\
Layer 1 (Institutional) & 4 & 84 & Mixed & 0.72 & Institutional indicators \\
Layer 2 (Firms) & 82 & 84 & Cross-sectional rank & 0.60 & Investment ratios \\
\midrule
\quad Diversified & 23 & & & & \\
\quad Technology & 15 & & & & \\
\quad Energy & 12 & & & & \\
\quad Financials & 10 & & & & \\
\quad Industrials & 12 & & & & \\
\quad Healthcare & 10 & & & & \\
\midrule
\textbf{Total} & \textbf{93} & \textbf{84} & & & \\
\bottomrule
\end{tabular}
\end{table}

\subsection{Validation Panels}

Two additional panels test the scope of the findings:

\begin{itemize}[nosep]
  \item \textbf{146-firm CapEx/Revenue:} Homogeneous panel---all firms, one ratio type, cross-sectional percentile ranks. No layer heterogeneity.
  \item \textbf{270-actor multi-ratio:} 135 firms $\times$ 2 ratios (CapEx/Revenue and Revenue/Assets). Moderate heterogeneity from ratio-type mixing, but no macro/institutional actors.
\end{itemize}

\subsubsection*{UK/EU Heterogeneous Panel and Combined Panel}
\label{sec:uk_eu_panel_intro}

Section~\ref{sec:cross_regime} introduces additional validation panels: a 109-actor UK/EU heterogeneous panel (Layer~0: 6 macro shocks from FRED EU proxies and BoE Bank Rate; Layer~1: 3 institutional intermediaries---ECB, BoE, IMF; Layer~2: 100 EU/UK-domiciled firms across nine European exchanges) sampled 2011Q2--2025Q4, and a combined US + UK/EU panel of 202 actors over the same 59-quarter overlap window. Both panels use the cross-sectional CapEx/Assets percentile-rank construction. The UK/EU firm universe is restricted to firms with EU/UK ISIN prefixes; detailed construction in Appendix~\ref{app:uk_eu}.

\subsection{Evaluation Protocol}

All results use 10 rolling out-of-sample windows (test years 2015--2024). Each window uses a 5-year training period ending at the start of the test year. Models are re-estimated quarterly within each test year as new observations arrive, so the training set expands from 20 to 24 quarters during each test year. The primary metric is out-of-sample $R^2$, defined as $R^2 = 1 - \sum_{i,t}(y_{i,t} - \hat{y}_{i,t})^2 / \sum_{i,t}(y_{i,t} - \bar{y}^{\mathrm{test}}_i)^2$, where $\bar{y}^{\mathrm{test}}_i$ is the test-window mean for actor $i$, computed from the four quarterly test observations within each test year. This is the standard in-sample $R^2$ formula applied to held-out data (equivalent to regressing actuals on predictions over the test set). The denominator uses test-set information; the metric therefore measures explained variance \emph{within} each test window rather than improvement over a training-set baseline. Because all model comparisons (Tables~\ref{tab:five_arch}--\ref{tab:cross_panel}) use the same denominator for both the numerator and reference model, the \emph{sign and significance} of the $\Delta R^2$ differentials are unaffected by this choice; the numerical magnitude of $\Delta$ does shift across conventions (Section~\ref{sec:validation} reports $\Delta = +0.067$ under the \citealt{stock2002forecasting} training-mean convention versus $+0.047$ under the test-window convention, because the denominator scalar differs). Statistical comparisons use paired-window percentile bootstrap confidence intervals (10{,}000 resamples; with $n = 10$ windows resampled with replacement, $\binom{19}{10} = 92{,}378$ distinct resamples exist, so bootstrap $p$-values cannot be finer than ${\sim}10^{-4}$) and modified Diebold--Mariano $t$-tests \citep{diebold1995comparing,harvey1997testing}. The Harvey--Leybourne--Newbold small-sample correction \citep{harvey1997testing} produces a multiplicative adjustment of $\sqrt{(n + 1 - 2 \cdot 1) / n} = \sqrt{0.9} \approx 0.949$ at $n = 10$, $h = 1$; this is negligible and does not change any significance conclusion. The placebo benchmark uses 1{,}000 random block partitions of identical sizes.

\textbf{Contextualisation of $R^2$ levels.} The prediction targets are cross-sectional percentile ranks, which are bounded $[0,1]$, highly autocorrelated ($\rho \approx 0.60$), and compressed toward the centre. Per-actor AR(1) already achieves $R^2 = 0.594$; the reported absolute $R^2$ levels are therefore not analogous to $R^2$ on returns or growth rates. The contribution of this paper is the \emph{differential} $\Delta R^2 = +0.047$ between architectures, not the absolute level. Mean absolute error (MAE) confirms the finding: M2 reduces MAE from 0.129 to 0.120 versus G1 (10/10 windows, $t = 7.93$).

\section{The Two-Stage Architecture}
\label{sec:architecture}

\subsection{Stage 1: Global Pooled AR(1) with Fixed Effects}

The first stage estimates shared persistence from the full panel:
\begin{equation}
\hat{y}^{\mathrm{pool}}_{i,t+1} = \bar{y}_i + \hat{\rho} \left( y_{i,t} - \bar{y}_i \right)
\end{equation}
where $\bar{y}_i$ is the actor-specific mean (fixed effect) and $\hat{\rho}$ is the pooled persistence parameter, both estimated by within-transformation OLS on the training data only and re-estimated quarterly as new observations arrive. Global estimation benefits from the larger effective sample for a single pooled $\hat\rho$. While persistence varies across layers (Table~\ref{tab:panel_stats}: macro $\rho \approx 0.88$, firms $\rho \approx 0.60$), the pooled estimator captures an effective average persistence; the layer-specific residual structure is then handled by Stage~2. Whether block-specific $\hat\rho_b$ would further improve Stage~1 is tested directly: the BA\_M2 architecture (Table~\ref{tab:five_arch}) combines block-specific $\hat\rho_b$ with block-specific Stage~2 and achieves $R^2 = 0.661$---significantly \emph{worse} than M2's $0.677$ ($t = 3.18$, $p = 0.011$), demonstrating that the global pooled $\hat\rho$ actually helps the local Stage~2 by retaining cross-block information in the residuals.

\subsection{Stage 2: Residual Dynamics}

The second stage operates on Stage~1 residuals $r_{i,t} = y_{i,t} - \hat{y}^{\mathrm{pool}}_{i,t}$, exponentially-weighted demeaned with a half-life of 12~quarters (3~years). These residuals capture the cross-sectional structure that shared persistence does not explain.

Three interchangeable engines model the residual dynamics:
\begin{enumerate}[nosep]
  \item \textbf{PCA + diagonal AR:} Extract $K$ principal components from the exponentially-weighted residual covariance, fit per-component AR(1) coefficients.
  \item \textbf{DMD + reduced operator:} Dynamic Mode Decomposition provides a rank-$K$ approximation of the one-step propagator via the Koopman framework \citep[Appendix~A;][]{roshka2026spectral}. The diagonal or full reduced propagator $\tilde{A}$ serves as the transition matrix.
  \item \textbf{Ridge regression:} Direct regularised regression of $r_{t+1}$ on $r_t$ in the full $N$-dimensional space.
\end{enumerate}

The combined forecast is $\hat{y}_{i,t+1} = \hat{y}^{\mathrm{pool}}_{i,t+1} + \hat{r}_{i,t+1}$.

\subsection{Why Two Stages?}

A standalone spectral model (DMD-based state-space filter with $K = 8$ modes) achieves $R^2 = 0.486$---far below per-actor AR(1) at 0.594. The single spectral basis pools heterogeneous actors, losing actor-specific persistence that AR(1) captures by construction. The two-stage architecture resolves this: Stage~1 handles persistence; Stage~2 handles the cross-sectional rotational structure in the residual. The transition diagnostic is detailed in Appendix~B.

Table~\ref{tab:augmentation} reports the augmentation gain on all three panels. The gain is robust across panels with different compositions, ranging from $+1.7$~pp on the homogeneous 146-firm panel to $+3.6$~pp on the heterogeneous 93-actor panel.

\begin{table}[ht]
\centering
\caption{Two-stage augmentation vs baselines. $\Delta$ is the augmentation gain (augmented $R^2$ minus AR(1) $R^2$). All three panel-average gains are positive (the 93-actor panel CI and per-window detail are reported in Table~\ref{tab:five_arch}).}
\label{tab:augmentation}
\small
\begin{tabular}{lcccc}
\toprule
Panel & $N$ & AR(1) $R^2$ & Augmented $R^2$ & Gain $\Delta$ \\
\midrule
93-actor multilayer & 93 & 0.594 & 0.630 & $+0.036$ \\
146-firm CapEx/Revenue & 146 & 0.728 & 0.745 & $+0.017$ \\
270-actor multi-ratio & 270 & 0.728 & 0.753 & $+0.025$ \\
\bottomrule
\end{tabular}
\end{table}

Table~\ref{tab:augmentation} reports the gain of global augmentation (G1) over per-actor AR(1). The additional gain from the mixture architecture over G1 is reported separately in Table~\ref{tab:five_arch}.

\subsection{Method Equivalence}
\label{sec:method_equiv}

Table~\ref{tab:method_comparison} compares 12~models across four complexity classes on the 93-actor panel. At matched effective complexity, PCA, DMD, and Ridge achieve statistically indistinguishable predictive $R^2$:

\begin{table}[H]
\centering
\caption{Method comparison: 12 models on the 93-actor panel. All models use global pooled+FE as Stage~1 and are re-estimated quarterly. $\Delta$ is versus rolling per-actor AR(1) ($R^2 = 0.610$), which is also re-estimated quarterly for a fair comparison (the fixed-parameter AR(1) in Table~\ref{tab:augmentation} achieves $R^2 = 0.594$; the rolling version is slightly better). Forecast-error correlations $\rho_{\mathrm{pred}}$ confirm predictions are functionally identical, not just equally accurate on average.}
\label{tab:method_comparison}
\small
\begin{tabular}{llccccl}
\toprule
Class & Model & $R^2$ & $\Delta$ roll.\ AR(1) & $t$ & $p$ & CI \\
\midrule
\multirow{4}{*}{Tiny (${\sim}K$)} & PCA+diag Kalman & 0.622 & $+0.012$ & 1.97 & 0.080 & $[+0.002, +0.024]$ \\
& DMD+diag Kalman & 0.619 & $+0.009$ & 2.07 & 0.068 & $[+0.001, +0.017]$ \\
& PCA reduced (no Kalman) & 0.621 & $+0.011$ & 1.78 & 0.109 & $[+0.001, +0.023]$ \\
& DMD reduced (no Kalman) & 0.618 & $+0.008$ & 1.86 & 0.096 & $[+0.000, +0.016]$ \\
\midrule
\multirow{4}{*}{Medium (${\sim}K^2$)} & PCA+full VAR & 0.577 & $-0.033$ & $-3.86$ & 0.004 & $[-0.049, -0.017]$ \\
& DMD+full $\tilde{A}$ Kalman & 0.630 & $+0.020$ & 4.22 & 0.002 & $[+0.012, +0.029]$ \\
& PCA+ridge VAR & 0.619 & $+0.009$ & 1.84 & 0.099 & $[-0.000, +0.019]$ \\
& Reduced-rank Ridge $K{=}8$ & 0.629 & $+0.019$ & 3.19 & 0.011 & $[+0.009, +0.031]$ \\
\midrule
Large (reg.\ $N \!\times\! N$) & Ridge on residuals & 0.632 & $+0.022$ & 3.84 & 0.004 & $[+0.012, +0.033]$ \\
\midrule
\multirow{3}{*}{Non-linear} & GBM (per-actor, global) & 0.661 & $+0.051$ & 2.94 & 0.016 & $[+0.017, +0.080]$ \\
& GBM (per-actor, block-specific) & 0.657 & $+0.047$ & 2.78 & 0.021 & $[+0.015, +0.077]$ \\
& GBM + sector features & 0.592 & $-0.018$ & $-0.81$ & 0.440 & $[-0.060, +0.022]$ \\
\bottomrule
\end{tabular}

\vspace{0.3em}
{\footnotesize Forecast-error correlations (linear models): $\rho(\text{DMD}, \text{PCA}) = 0.990$; $\rho(\text{DMD}, \text{Ridge}) = 0.980$; $\rho(\text{PCA}, \text{Ridge}) = 0.969$.\\
Under Holm--Bonferroni adjustment for 12 comparisons, three models remain significant at the 5\% level. Two are significantly \emph{better} than rolling AR(1): DMD+full $\tilde{A}$ Kalman ($+0.020$) and Ridge ($+0.022$). One is significantly \emph{worse}: PCA+full VAR ($-0.033$), which overfits at $K^2 = 64$ unrestricted parameters. GBM global ($+0.051$, $p = 0.016$), GBM block-specific ($+0.047$, $p = 0.021$), and reduced-rank Ridge ($+0.019$, $p = 0.011$) are individually significant but do not survive the Holm step-down at ranks 4--6. The remaining models---including GBM with sector features ($-0.018$, n.s.)---are non-significant, \emph{consistent with} the method-equivalence claim at matched effective complexity among linear estimators.}
\end{table}

Among the matched-complexity linear reduced-rank estimators tested, the forecasting ceiling is $R^2 \approx 0.630$. Per-actor GBM on Stage~1 residuals breaks through this linear ceiling ($R^2 = 0.661$) but remains below M2 ($R^2 = 0.677$, Table~\ref{tab:five_arch}), confirming the ceiling is architectural rather than a linear-method artefact. Applying the same block decomposition to GBM---training separate per-actor GBMs using only within-block residuals as features for local blocks and full residuals for the remainder---does not help: block-specific GBM achieves $R^2 = 0.657$, slightly \emph{below} global GBM ($-0.004$, n.s.) and well below M2 ($-0.020$, $t = -3.05$, $p = 0.014$, 1/10 windows positive). Restricting GBM's feature set to within-block residuals removes cross-block information that the non-linear model can use without interference, unlike PCA which pools heterogeneous dynamics into a shared basis. The M2 gain therefore has two components: the architectural block decomposition \emph{and} the low-rank PCA+ridge representation, which extracts within-block factor structure that per-actor GBM---even with the same block decomposition---cannot capture. Equipping a single pooled GBM with sector/layer categorical features and sector$\times$residual interaction terms degrades performance further to $R^2 = 0.592$ (0/10 windows positive vs G1), because the pooled non-linear model overfits the high-dimensional interaction space at this sample size.

The equivalence holds at matched effective complexity among small-$K$ models: DMD with $K = 8$ diagonal parameters achieves $R^2 = 0.619$, while Ridge (heavily regularised $N \!\times\! N$) achieves $R^2 = 0.632$. That an 8-parameter spectral model nearly matches a regularised regression with orders-of-magnitude more effective parameters suggests the exploitable residual structure is genuinely low-rank. The practical implication: in settings where parsimony or interpretability matters, the spectral representation offers comparable predictive performance at a fraction of the complexity. Note that the equivalence is not universal: PCA with an unrestricted $K \!\times\! K$ VAR ($R^2 = 0.577$, significantly \emph{worse} than AR(1)) diverges from DMD with the full reduced operator ($R^2 = 0.630$), because the PCA VAR is not regularised and overfits at $K^2 = 64$ parameters on $T \approx 20$ quarters.

This method-equivalence result at the global level---together with the finding that PCA+ridge dominates DMD within individual blocks (Section~\ref{sec:falsification}), where smaller samples penalise the less regularised estimator---motivates the architectural focus of the remainder of the paper.

\subsection{The Heterogeneity-Aware Mixture Architecture}
\label{sec:mixture_arch}

The method-equivalence finding raises the question: if the engine does not matter, what \emph{does} matter? The answer, developed in Section~\ref{sec:results}, is the \emph{granularity of the decomposition}. The mixture architecture replaces the single global Stage~2 with block-specific local models:

\begin{enumerate}[nosep]
  \item \textbf{Stage 1} (unchanged): Global pooled AR(1)+FE on all 93 actors.
  \item \textbf{Block assignment}: Each actor is assigned to a block based on economic sector and data-type structure (Table~\ref{tab:blocks}).
  \item \textbf{Stage 2} (block-specific): For each local block, estimate a local PCA+ridge model on within-block Stage~1 residuals. For the remainder block (actors well-served by the global model), retain the global augmentation.
\end{enumerate}

\begin{table}[H]
\centering
\caption{Selected block assignments and economic rationales. Blocks were chosen from 10 candidates based on per-block $\Delta R^2$ and window consistency (Appendix~\ref{app:candidates}).}
\label{tab:blocks}
\small
\begin{tabular}{lccl}
\toprule
Block & $N$ & Local $K$ & Economic rationale \\
\midrule
Diversified & 23 & 4 & Heterogeneous sector; global basis misrepresents \\
Macro/Institutional & 11 & 2 & Different data types (FRED + institutional) \\
Tech/Health & 25 & 4 & Coherent sub-panel (high within-block loading similarity) \\
Remainder & 34 & --- & Energy, industrials, financials (global model adequate) \\
\bottomrule
\end{tabular}
\end{table}

Block assignments are fixed before execution of the rolling evaluation protocol using static economic metadata from the actor registry. The selection of which blocks to treat locally was informed by preliminary per-block diagnostics during exploratory analysis (Section~\ref{sec:discussion}, Appendix~\ref{app:candidates}).

\begin{figure}[ht]
\centering
\begin{tikzpicture}[
  box/.style={rectangle, draw=accent!60, thick, rounded corners=3pt,
              align=center, inner sep=6pt, font=\small},
  sbox/.style={rectangle, draw=accent!40, rounded corners=2pt,
              align=center, inner sep=5pt, fill=accent!3, font=\small},
  lbox/.style={rectangle, draw=red!50, thick, rounded corners=3pt,
              align=center, inner sep=5pt, fill=red!4, font=\small},
  arr/.style={-{Stealth[length=2.5mm]}, thick, color=accent!60},
  larr/.style={-{Stealth[length=2.5mm]}, thick, color=red!50},
  node distance=0.55cm
]
\node[font=\small\bfseries, text=accent] at (-3.8, 4.0) {(a) Global};
\node[box, fill=accent!6, minimum width=5.5cm, text width=5cm] (inp1) at (-3.8, 3.2)
  {\textbf{Panel} $Y \in [0,1]^{93 \times T}$};
\node[box, fill=blue!5, minimum width=5.5cm, text width=5cm, below=of inp1] (s1a)
  {\textbf{Stage 1:} Global pooled AR(1)+FE\\[-1pt]
   {\footnotesize $\hat{\rho}$ from all 93 actors}};
\node[box, fill=orange!5, minimum width=5.5cm, text width=5cm, below=of s1a] (s2a)
  {\textbf{Stage 2:} Global residual model\\[-1pt]
   {\footnotesize One basis for all actors ($K{=}8$)}};
\node[sbox, minimum width=5.5cm, below=of s2a] (out1)
  {$\hat{y} = \hat{y}^{\mathrm{pool}} + \hat{r}^{\mathrm{global}}$\\[-1pt]
   {\footnotesize $R^2 = 0.630$}};
\draw[arr] (inp1) -- (s1a);
\draw[arr] (s1a) -- (s2a);
\draw[arr] (s2a) -- (out1);

\node[font=\small\bfseries, text=red!60!black] at (3.8, 4.0) {(b) Mixture};
\node[box, fill=accent!6, minimum width=5.5cm, text width=5cm] (inp2) at (3.8, 3.2)
  {\textbf{Panel} $Y \in [0,1]^{93 \times T}$};
\node[box, fill=blue!5, minimum width=5.5cm, text width=5cm, below=of inp2] (s1b)
  {\textbf{Stage 1:} Global pooled AR(1)+FE\\[-1pt]
   {\footnotesize $\hat{\rho}$ from all 93 actors (same)}};
\node[lbox, minimum width=5.5cm, text width=5cm, below=of s1b] (s2b)
  {\textbf{Stage 2:} Block-specific local models\\[-1pt]
   {\footnotesize Diversified: local PCA+ridge ($K{=}4$)}\\[-1pt]
   {\footnotesize Tech/Health: local PCA+ridge ($K{=}4$)}\\[-1pt]
   {\footnotesize Macro/Inst: local PCA+ridge ($K{=}2$)}\\[-1pt]
   {\footnotesize Remainder: global augmentation}};
\node[sbox, minimum width=5.5cm, below=0.45cm of s2b] (out2)
  {$\hat{y}_i = \hat{y}^{\mathrm{pool}}_i + \hat{r}^{\mathrm{local}}_{b(i)}$\\[-1pt]
   {\footnotesize $R^2 = 0.677$}};
\draw[arr] (inp2) -- (s1b);
\draw[larr] (s1b) -- (s2b);
\draw[larr] (s2b) -- (out2);
\end{tikzpicture}
\caption{Architecture comparison. (a)~The global pipeline applies one residual model to all 93~actors. (b)~The mixture pipeline uses the same global Stage~1 but routes actors to block-specific local models in Stage~2, with economically motivated blocks. Stage~1 is identical; only the Stage~2 routing differs.}
\label{fig:pipeline}
\end{figure}
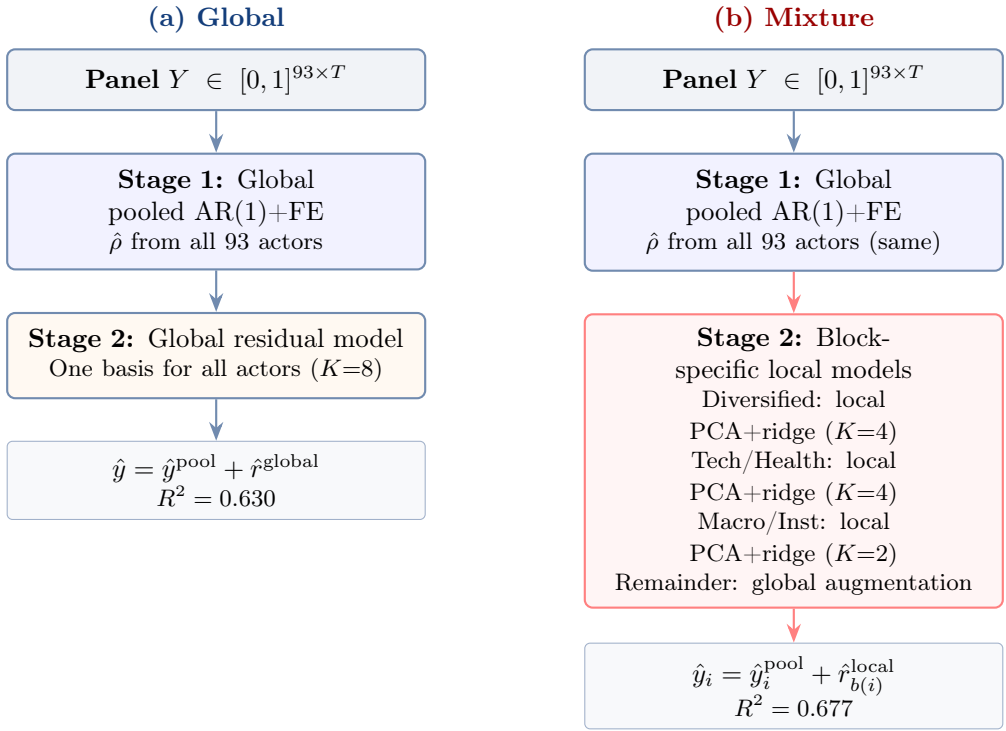

Formally, the mixture estimator is:
\begin{equation}
\hat{y}_{i,t+1} = \bar{y}_i + \hat{\rho}\,(y_{i,t} - \bar{y}_i) +
\begin{cases}
\bigl[\hat{U}_b \hat{A}_b \hat{U}_b^\top\, \mathbf{r}_{b,t}\bigr]_i & \text{if } b(i) \in \mathcal{B}_{\mathrm{local}} \\[3pt]
\bigl[\hat{U}_{\mathrm{g}} \hat{A}_{\mathrm{g}} \hat{U}_{\mathrm{g}}^\top\, \mathbf{r}_t\bigr]_i & \text{if } b(i) = \text{remainder}
\end{cases}
\label{eq:mixture}
\end{equation}
where $\hat{\rho}$ is pooled across all actors (global Stage~1; the homogeneity restriction applies to $\rho$ only, not to the factor loadings), $\bar{y}_i$ is the actor fixed effect, $b(i)$ is the assigned block, $\mathcal{B}_{\mathrm{local}} = \{\text{Diversified, Macro/Inst, Tech/Health}\}$ are the locally-treated blocks, $\mathbf{r}_{b,t} \in \R^{N_b}$ is the vector of Stage~1 residuals for actors in block $b$, $\hat{U}_b \in \R^{N_b \times K_b}$ and $\hat{A}_b \in \R^{K_b \times K_b}$ are the block-specific PCA basis and ridge VAR, $[\cdot]_i$ selects the component corresponding to actor $i$'s position within its block (i.e., if actor $i$ is the $j$-th member of block $b$, $[\cdot]_i \equiv [\cdot]_j$ in the local $N_b$-dimensional space), and $\hat{U}_{\mathrm{g}}$, $\hat{A}_{\mathrm{g}}$ are the global basis and operator applied to the full residual vector $\mathbf{r}_t \in \R^N$ for remainder actors. This is analogous to a known-group interactive fixed-effects model \citep{bai2009panel} with block-specific factor loadings. Setting $\hat{U}_b = 0$ for all local blocks and $\hat{U}_{\mathrm{g}} = 0$ recovers pooled-only (G0) exactly. The global augmentation (G1) applies $\hat{U}_{\mathrm{g}}, \hat{A}_{\mathrm{g}}$ to all actors; the mixture (M2) replaces this with block-specific operators for actors in $\mathcal{B}_{\mathrm{local}}$. Figure~\ref{fig:pipeline} contrasts the two pipelines.

\section{When Global Models Misalign}
\label{sec:results}

\subsection{Full-Panel Result}

Eight architectures are compared. The core five: G0~(pooled-only, no augmentation), G1~(global always-on augmentation), S1~(selective-off: pooled-only for local blocks, global augmentation for the remainder---a diagnostic that isolates harm-removal from local modelling), M1~(mixture with local Ridge), M2~(mixture with local PCA+ridge). Three additional baselines address specific alternative explanations: BA~(block-specific $\hat\rho_b$+FE, no Stage~2), BA\_M2~(block-specific $\hat\rho_b$ + block-specific Stage~2, testing whether the M2 gain is from correcting Stage~1 persistence misspecification), and ENS~(equal-weighted ensemble of G1 and BA, testing whether simple forecast combination replicates the mixture gain).

\begin{figure}[ht]
\centering
\includegraphics[width=0.85\textwidth]{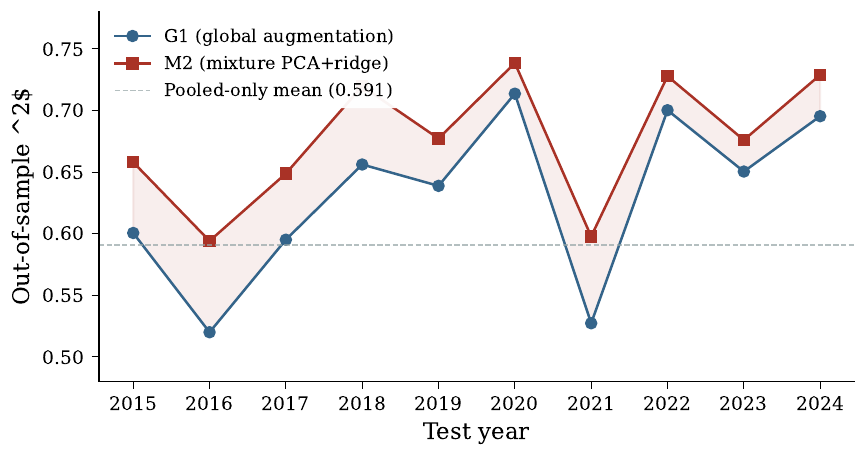}
\caption{Out-of-sample $R^2$ by test year for global always-on augmentation (G1) and the mixture architecture (M2). M2 exceeds G1 in all 10 windows. Horizontal reference lines show pooled-only and per-actor AR(1) baselines.}
\label{fig:per_window}
\end{figure}

Table~\ref{tab:five_arch} reports all eight architectures on the full 93-actor panel; Figure~\ref{fig:per_window} shows the per-window comparison. The mixture architectures (M1, M2) beat the global always-on augmentation (G1) in every window.

\begin{table}[H]
\centering
\caption{Eight architectures on the full 93-actor panel (10 rolling OOS windows, 2015--2024). S1~applies pooled-only (no augmentation) to local blocks and global augmentation to the remainder. M1 and M2 apply local Ridge or local PCA+ridge to local blocks. BA\_M2 combines block-specific $\hat\rho_b$ (Stage~1) with block-specific PCA+ridge (Stage~2). ENS is the equal-weighted average of G1 and BA predictions. W denotes the number of windows (out of 10) with a positive gain over G1.}
\label{tab:five_arch}
\small
\begin{tabular}{llcccccc}
\toprule
\# & Architecture & $R^2$ & $\Delta$ vs G1 & $t$ & $p$ & CI & W \\
\midrule
G0 & Pooled-only & 0.591 & $-0.039$ & $-7.14$ & ${<}0.001$ & $[-0.048, -0.029]$ & 0/10 \\
BA & Block-specific $\hat\rho_b$+FE & 0.611 & $-0.019$ & $-3.97$ & $0.003$ & $[-0.029, -0.011]$ & 0/10 \\
G1 & Global always-on & 0.630 & --- & --- & --- & --- & --- \\
ENS & Ensemble(G1, BA) & 0.639 & $+0.009$ & $5.28$ & ${<}0.001$ & $[+0.006, +0.012]$ & 9/10 \\
S1 & Selective-off & 0.599 & $-0.031$ & $-8.55$ & ${<}0.001$ & $[-0.037, -0.024]$ & 0/10 \\
BA\_M2 & Block $\hat\rho_b$ + local S2 & 0.661 & $+0.031$ & $4.83$ & ${<}0.001$ & $[+0.019, +0.044]$ & 10/10 \\
\textbf{M1} & \textbf{Mixture (Ridge)} & \textbf{0.669} & $\mathbf{+0.039}$ & $\mathbf{7.80}$ & $\mathbf{{<}0.001}$ & $\mathbf{[+0.030, +0.049]}$ & \textbf{10/10} \\
\textbf{M2} & \textbf{Mixture (PCA+ridge)} & \textbf{0.677} & $\mathbf{+0.047}$ & $\mathbf{7.76}$ & $\mathbf{{<}0.001}$ & $\mathbf{[+0.036, +0.058]}$ & \textbf{10/10} \\
\bottomrule
\end{tabular}
\end{table}

Under Holm--Bonferroni correction for the 7 comparisons against G1, M2 ($p < 0.001$, threshold $0.05/7 = 0.0071$), M1 ($p < 0.001$), BA\_M2 ($p < 0.001$), and ENS ($p < 0.001$) all remain significant at the 5\% level; S1, G0, and BA (all negative) likewise remain significant. No significance conclusion changes.

The block-specific $\hat\rho_b$+FE baseline (BA) --- which allows each block its own persistence parameter but has no Stage~2 --- achieves $R^2 = 0.611$, well below M2 at 0.677. The interacting combination BA\_M2 --- block-specific $\hat\rho_b$ \emph{combined with} block-specific Stage~2 --- achieves $R^2 = 0.661$ ($\Delta = +0.031$, 10/10 windows). This is significantly better than BA alone ($+0.050$) but \emph{worse} than M2 ($-0.016$, $t = 3.18$, $p = 0.011$, CI $[+0.008, +0.026]$, 9/10 windows positive), demonstrating that the global pooled $\hat\rho$ in Stage~1 actually \emph{helps} the local Stage~2: the structured persistence residuals from global pooling contain cross-block information that block-specific $\hat\rho_b$ discards. The M2 gain is genuine local residual dynamics, not a correction for Stage~1 misspecification.

The selective-off architecture (S1) is \emph{worse} than global ($\Delta = -0.031$), demonstrating that the mixture gain is not from removing suboptimal augmentation. The local models genuinely add predictive value: M1 exceeds S1 by $+0.070$ ($t = 16.12$, $p < 0.001$).

The primary comparison (M2 vs G1) is robust to dependence correction. With Newey--West HAC variance (bandwidth 1--3), the DM-HAC statistic ranges from 6.84 to 7.38 ($p < 0.001$ in all cases). A moving-block bootstrap (block lengths 2 and 3) yields CIs of $[+0.035, +0.059]$ and $[+0.036, +0.059]$, both excluding zero. All DM tests use $n = 10$ windows ($\mathrm{df} = 9$); at these degrees of freedom, HAC bandwidth 1--3 consumes 10--30\% of the sample in kernel weighting, and bootstrap estimators have limited resampling resolution ($\binom{10+10-1}{10} = 92{,}378$ distinct paired resamples). The DM-HAC statistics are therefore best interpreted as a sanity check on the primary paired bootstrap, not as independent asymptotic inference \citep[cf.][for finite-sample bias in HAC estimators]{kiefer2005new}. The primary inferential weight rests on the bootstrap CIs and the exact sign test. Nonetheless, the conclusion is invariant to the treatment of window-level dependence. A non-parametric sign test confirms the finding: M2 exceeds G1 in all 10 windows (exact binomial $p = 2^{-10} \approx 0.001$ under the null of equal performance). Because the four within-year forecasts share an expanding training set, the 10 annual windows are not fully independent. The lag-1 autocorrelation of the window-level $R^2$ differentials $(d_w = R^2_{\mathrm{M2},w} - R^2_{\mathrm{G1},w})$ is $\hat\rho_d = 0.11$, yielding an effective sample size $n_{\mathrm{eff}} = n \cdot (1 - \hat\rho_d)/(1 + \hat\rho_d) \approx 10 \cdot 0.89/1.11 \approx 8.0$. With $n_{\mathrm{eff}} \approx 8$ effective independent blocks, the sign-test $p$-value rises to $2^{-8} \approx 0.004$, which remains significant.\footnote{M2 and G1 are not nested in the standard sense: M2 replaces the global Stage~2 with block-specific models estimated on different data subsets, rather than adding parameters to G1. The \citet{clark2007approximately} adjustment for nested forecast comparisons therefore does not directly apply to the headline M2-vs-G1 comparison; the \citet{giacomini2006tests} framework for non-nested comparisons under rolling re-estimation is the more appropriate reference. For nested pairs (G0$\subset$G1, S1$\subset$M1), the standard DM statistic is conservative (biased toward the null), so the reported significance is, if anything, understated.}

A direct forecast combination \citep{bates1969combination} does not replicate the mixture gain. The equal-weighted ensemble of G1 (global augmentation) and BA (block-specific $\hat\rho_b$+FE) achieves $R^2 = 0.639$ ($\Delta = +0.009$ vs G1, 9/10 windows; Table~\ref{tab:five_arch}). This is a modest improvement over G1 but far below M2's $+0.047$. The mixture architecture captures block-specific factor structure ($\hat{U}_b$, $\hat{A}_b$) estimated from within-block residuals---information not contained in any weighted average of global-basis and per-actor AR(1) forecasts. The information sets differ structurally, not just in weighting. (The ensemble's modest gain over G1 despite BA losing in all 10 windows reflects the standard Bates--Granger diversification effect: the squared-error correlation between G1 and BA forecasts is $\rho \approx 0.865$, and the roughly 13\% uncorrelated error mass suffices for averaging to reduce variance even when one input is uniformly worse.)

\subsection{Per-Block Decomposition}

\begin{figure}[ht]
\centering
\includegraphics[width=0.85\textwidth]{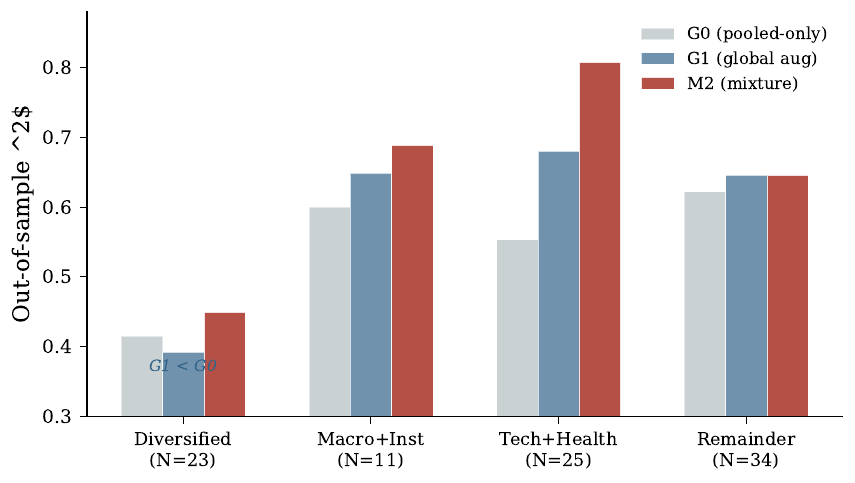}
\caption{Per-block $R^2$ under three architectures: pooled-only (G0), global augmentation (G1), and mixture PCA+ridge (M2). The diversified sector is degraded by global augmentation ($\text{G1} < \text{G0}$); the tech/health block benefits most from local treatment.}
\label{fig:per_block}
\end{figure}

Table~\ref{tab:per_block} and Figure~\ref{fig:per_block} reveal the mechanism. The global basis causes cross-block interference that manifests differently across blocks.

\begin{table}[H]
\centering
\caption{Per-block $R^2$ by architecture. The diversified sector is degraded by global augmentation (0.392~$<$~0.415 pooled-only). The tech/health block gains most from local treatment. Note: per-block $R^2$ uses the test-window mean $\bar{y}^{\mathrm{test}}_i$ in the denominator (consistent with the definition in Section~\ref{sec:data}), summing only over actors $i$ in that block, so the $N$-weighted average of per-block $R^2$'s does not equal the full-panel $R^2$ in Table~\ref{tab:five_arch}.}
\label{tab:per_block}
\small
\begin{tabular}{lcccccc}
\toprule
Block & $N$ & G0 (pooled) & G1 (global) & S1 (sel-off) & M1 (Ridge) & M2 (PCA+r) \\
\midrule
Diversified & 23 & 0.415 & 0.392 & 0.415 & 0.461 & 0.449 \\
Macro/Institutional & 11 & 0.600 & 0.649 & 0.600 & 0.692 & 0.689 \\
Tech/Health & 25 & 0.554 & 0.681 & 0.554 & 0.764 & 0.808 \\
Remainder & 34 & 0.622 & 0.646 & 0.646 & 0.646 & 0.646 \\
\bottomrule
\end{tabular}
\end{table}

Three patterns emerge. First, the diversified sector is the clearest case of cross-block interference: global augmentation \emph{reduces} $R^2$ from 0.415 (pooled-only) to 0.392. The global basis, dominated by technology/healthcare and macro modes, injects noise into predictions for this heterogeneous sector. Local Ridge recovers $R^2 = 0.461$.

Second, the tech/health block shows the largest absolute gain. Global augmentation already helps substantially ($+0.127$ over pooled), but local PCA+ridge with $K = 4$ nearly doubles the gain to $R^2 = 0.808$. The within-block co-movement structure is captured far more efficiently by a local 4-component model than by a global 8-component model that mixes in macro and diversified dynamics.

Third, the remainder block (energy, industrials, financials) is unchanged by design---the same global augmentation is applied in all architectures. The global basis serves these sectors well; local decomposition is not needed.

\subsection{Geometric Description}
\label{sec:geodesic}

At the rank $K{=}4$ used for block-level comparison (block sizes preclude the model's operating rank $K{=}8$; the $K{=}8$ rotation and its unpredictability are analysed in Section~\ref{sec:falsification}), the global spectral basis rotates at $31.5^\circ$ per quarter (geodesic distance on the Grassmannian); within-block rotation ranges from $14^\circ$ to $28^\circ$ (per-block values available in the replication archive). For reference, two uniformly random $K$-dimensional subspaces in $\R^N$ have expected geodesic distance $K \cdot \arccos(0) = K \cdot 90^\circ / \sqrt{\pi K/2} \approx 63^\circ$ at $K{=}4$, $N{=}93$; the observed $31.5^\circ$ is well below this random baseline, confirming the basis has genuine temporal persistence. However, a matched-size random sub-panel control (200 draws per block size, same $K$ and metric) shows the within-block reduction relative to the global rotation is \emph{not statistically significant}: Tech/Health $p = 0.080$; Macro/Institutional $p = 0.145$; Diversified $p = 0.305$. The lower within-block rotation is largely a mechanical effect of smaller panel size producing more stable estimated bases. The geodesic evidence is therefore descriptive context, not a demonstrated mechanism---the forecasting result rests on the actual $R^2$ comparison (Table~\ref{tab:five_arch}; placebo $z = 7.82$, Table~\ref{tab:placebo}), not on the geometric story.


\section{Validation}
\label{sec:validation}

\subsection{Placebo Test}
\label{sec:placebo}

\begin{figure}[ht]
\centering
\includegraphics[width=0.80\textwidth]{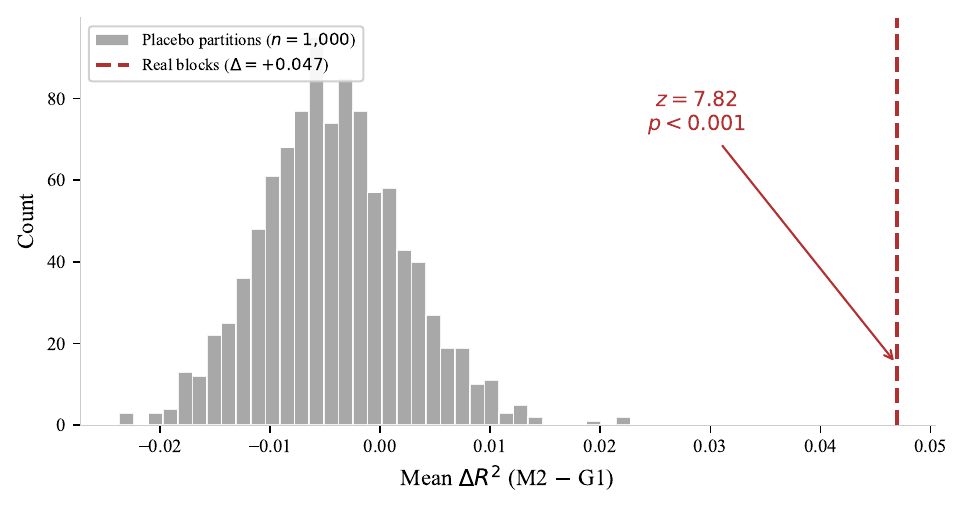}
\caption{Distribution of $\Delta R^2$ (mixture $-$ global) across 1{,}000 random block partitions of identical sizes (grey histogram) versus the real economic partition (vertical red line at $+0.047$). Random blocks produce a mean gain of $-0.004$; no random partition exceeds the real gain. The real gain exceeds the placebo distribution by $z = 7.82$ (Monte Carlo $p < 0.001$).}
\label{fig:placebo}
\end{figure}

The strongest potential objection to block-specific modelling is overfitting the block structure. We address this with a placebo benchmark (Table~\ref{tab:placebo}, Figure~\ref{fig:placebo}): 1{,}000 random partitions of the 93~actors into blocks of exactly the same sizes as the real partition (23, 11, 25, 34). For each random partition, the full mixture pipeline (M2) is run with identical local model specifications.

\begin{table}[H]
\centering
\caption{Placebo test: 1{,}000 random partitions vs real economic blocks.}
\label{tab:placebo}
\small
\begin{tabular}{lc}
\toprule
Statistic & Value \\
\midrule
Real economic blocks $\Delta$ & $+0.047$ \\
Placebo mean $\Delta$ & $-0.004$ \\
Placebo std & $0.0065$ \\
Placebo maximum & $+0.023$ \\
Placebo 99th percentile & $+0.012$ \\
$z$-score (real vs placebo) & $7.82$ \\
Placebo $p$-value & ${\le}\,0.001$ (0/1{,}000; Monte Carlo bound $1/1{,}001$) \\
\bottomrule
\end{tabular}
\end{table}

Random blocks degrade performance on average ($\Delta = -0.004$) because local estimation noise dominates when blocks are arbitrary. No random partition approaches the real gain. Each placebo uses the same 10 evaluation windows, same block sizes, and same local model specification---the only difference is actor assignment. The placebo benchmark controls for small-sample bias by design.

A natural question is whether concentration --- a random partition happening to place a coherent cluster in one block --- could produce a large aggregate gain. The evidence suggests not: the placebo maximum across 1{,}000 draws is $+0.023$, roughly half the real gain, and this includes any accidental coherence. Moreover, the real partition's gain is itself concentrated in the tech/health block (${\sim}72\%$; see Section~5.4), which means the placebo test is effectively comparing the real tech/health signal against the best accidental cluster in 1{,}000 random 25-actor sub-panels --- and the real signal clearly exceeds the best accidental cluster.

\subsection{Cross-Panel Validation}

\begin{table}[H]
\centering
\caption{Cross-panel validation across the original three test panels at training window $T=5$. The ``Heterogeneous?'' column refers to data-type heterogeneity (mixing macro/institutional series with firm-level ranks). Section~\ref{sec:ratio_robust} shows the 146-firm null is specific to the CapEx/Revenue ratio: a CapEx/Assets variant on the same 146 firms yields $\Delta = +0.018$ at $T=5$ (10/10 windows positive). Section~\ref{sec:cross_regime} extends the comparison to UK/EU and combined US+UK/EU panels.}
\label{tab:cross_panel}
\small
\begin{tabular}{lccccc}
\toprule
Panel & $N$ & Heterogeneous? & Global $R^2$ & Mixture $R^2$ & $\Delta$ \\
\midrule
\textbf{93-actor multilayer} & \textbf{93} & \textbf{Yes} & \textbf{0.630} & \textbf{0.677} & $\mathbf{+0.047}$ \\
146-firm CapEx/Revenue & 146 & No & 0.745 & 0.743 & $-0.003$ \\
270-actor multi-ratio & 270 & Moderate & 0.753 & 0.754 & $+0.001$ \\
\bottomrule
\end{tabular}
\end{table}

The mixture architecture is null on the 146-firm CapEx/Revenue panel and the 270-actor multi-ratio panel under the cross-sectional percentile-rank construction used for the multilayer panel.\footnote{A separate spectral architecture---standalone DMD+Kalman with $K{=}2$ operating directly on the EWM-demeaned panel without a pooled Stage~1---provides a small gain over per-actor AR(1) on the 146-firm panel ($+0.028$ $R^2$, 9/10 windows positive). This operates through a different mechanism (low-rank persistence compression rather than cross-block interference removal) and is the subject of a separate methodological investigation.} The 270-actor result reflects the ratio-type distinction: averaging the gain across CapEx/Revenue and Revenue/Assets cells dilutes any signal that arises in one but not the other.

\subsubsection*{Ratio-Choice Robustness on the 146-Firm Panel}
\label{sec:ratio_robust}

The 146-firm null reported above uses CapEx/Revenue as the cross-sectional ratio. A robustness check using CapEx/Assets on the same 146-firm population (top 146 firms by data completeness from the SEC EDGAR \texttt{PaymentsToAcquirePropertyPlantAndEquipment} $\div$ \texttt{Assets} ratio with the same $\geq 50\%$ non-NaN coverage filter) yields the opposite finding: $\Delta = +0.013$ at $T=3$ (paired-$t$ $p < 0.0001$, 10/10 windows positive) and $\Delta = +0.018$ at $T=5$ (paired-$t$ $p = 0.0002$, 10/10 windows positive), both with the same alphabetical-halves null partition used in Table~\ref{tab:cross_panel}. A larger 442-firm CapEx/Assets panel (same coverage filter, no top-146 subsetting) gives $\Delta = +0.020$ at both $T=3$ and $T=5$ ($p < 0.0001$, 10/10).

The single change---replacing Revenues with Assets in the denominator---turns the null into a positive significant gain even with the same null partition. The mixture architecture's gain depends on the cross-sectional dispersion of autoregressive structure across the panel actors. CapEx/Revenue ranks have lower autoregressive coherence: revenues respond to shorter-cycle business conditions; the ratio is volatile, so any partition contains comparable internal noise. CapEx/Assets ranks are more persistent (assets are slow-moving), and the cross-sectional ranks therefore retain block-coherent autoregressive structure that local PCA+ridge can exploit. The 270-actor multi-ratio null likely reflects dilution: half the actors carry the no-signal CapEx/Revenue ratio. Section~\ref{sec:cross_regime} confirms the refined scope condition holds cross-regime: a UK/EU firm-only panel using CapEx/Assets shows $\Delta = +0.015$ at $T=3$.

\subsection{Leave-One-Window-Out Block Selection}

A potential concern is that the choice of which blocks receive local treatment was informed by the same evaluation windows used to measure the gain. To address this, a strict leave-one-window-out (LOWO) procedure is applied: for each test year $w$, the block selection uses per-block $R^2$ from the \emph{other} 9 windows only. A block receives local treatment if $\bar\Delta_{-w} = \overline{R^2_{\text{local}} - R^2_{\text{global}}} > 0$ across the held-out windows.

All 10 windows select the same three blocks (Diversified, Macro/Institutional, Tech/Health). Since LOWO selects the identical partition every window, the resulting forecasts are the same as the fixed-partition mixture by construction---the gain is $+0.047$ with the same CI and significance.

An important qualification: LOWO controls for within-sample overfitting of the selection rule applied to the 10 candidate blocks, but it does not control for the researcher's prior choice to consider \emph{these} 10 candidates (6 raw sector blocks, 2 merged sector blocks, 2 layer blocks) rather than alternative candidate sets (e.g., size quantiles, loading-based clusters, geographic partitions). The researcher degree of freedom resides in the candidate-generation step, not the selection step. The placebo test (Section~\ref{sec:placebo}) controls for block \emph{sizes} but not the candidate space. A simpler two-block partition (Tech/Health local, remainder global) retains 66\% of the three-block gain ($\Delta = +0.031$, 10/10 windows), suggesting that even a minimally specified partition---one that requires selecting only the single most coherent sub-panel---captures the majority of the effect. Cross-panel replication (Table~\ref{tab:cross_panel}), which is immune to candidate-space search, provides the strongest external validation. The held-out decade test below provides a direct control for the candidate-space concern.

\subsection{Held-Out Decade Validation}
\label{sec:held_out}

To control for the candidate-generation degree of freedom, the block partition is frozen using only pre-2015 data. All 10 candidate blocks are evaluated on test years 2010--2014 (5 windows, 5-year training), and the selection rule $\Delta > 0$ and $W \geq 4/5$ is applied. The selected partition is then evaluated, without modification, on the unseen 2015--2024 decade.

Phase~A (2005--2014) selects 7 blocks: Diversified ($+0.018$, 5/5), Technology ($+0.007$, 4/5), Financials ($+0.001$, 4/5), Industrials ($+0.003$, 4/5), Tech/Health merged ($+0.034$, 5/5), Industrials/Energy ($+0.006$, 5/5), and Macro/Institutional ($+0.005$, 4/5). Healthcare ($W = 3/5$), Energy ($W = 3/5$), and All-Firms ($\Delta < 0$) are excluded.

Phase~B evaluates this frozen 7-block partition on 2015--2024: $\Delta = +0.050$ ($t = 9.11$, $p < 0.001$, CI $[+0.040, +0.061]$, 10/10 windows positive). The gain \emph{exceeds} the exploratory 3-block result ($+0.047$), confirming that the finding survives strict temporal separation of block selection and evaluation. The larger partition selected by Phase~A reflects the more aggressive selection rule (4/5 vs 7/10), but the core conclusion is unchanged: economically motivated blocks improve prediction on unseen data.

\subsection{Stratified Placebo}
\label{sec:strat_placebo}

The unstratified placebo (Section~\ref{sec:placebo}) allows random blocks to accidentally capture the Layer~0/1 vs Layer~2 data-type distinction, giving the real partition an unfair advantage on that dimension. A stratified placebo removes this advantage: the 11 Macro/Institutional actors (7 Layer~0 + 4 Layer~1) are fixed in their real block, and only the 82 firm-layer actors are randomly permuted across the remaining local blocks of sizes 23 (Diversified) and 25 (Tech/Health), with 34 actors assigned to the remainder (global augmentation). This is a strictly harder test: the placebo now controls for both block sizes \emph{and} the macro/firm data-type split.

Under this stratified design ($1{,}000$ permutations), the real partition achieves $\Delta = +0.047$; the stratified placebo mean is $+0.001$ (std $= 0.0063$, max $= +0.031$, 0/1{,}000 exceedances, $z = 7.25$, $p \leq 0.001$). The $z$-score drops from $7.82$ (unstratified) to $7.25$ (stratified), as expected---the data-type advantage accounts for roughly 7\% of the separation---but remains highly significant. The firm-sector assignment captures genuine within-Layer~2 co-movement structure that random firm partitions cannot replicate.

\subsection{Robustness: Dropping the Diversified Sector}

The diversified sector is the most heterogeneous block and the one where global augmentation most clearly degrades prediction. To test whether the headline gain depends on this specific block, the entire pipeline is re-estimated on the 70-actor panel after removing all diversified-sector actors. On this reduced panel (blocks: Macro/Institutional $N{=}11$, Tech/Health $N{=}25$, Remainder $N{=}34$), the mixture gain is $\Delta = +0.035$ ($t = 10.37$, CI $[+0.029, +0.042]$, 10/10 windows positive)---75\% of the original gain. The headline result does not depend primarily on the diversified sector: 75\% of the gain survives its removal, sustained primarily by the tech/health block. The remaining 25\% reflects the diversified block's specific contribution.\footnote{The diversified block's marginal contribution within the three-block mixture ($+0.012 = 0.047 - 0.035$) is slightly below its standalone single-block value in Appendix~\ref{app:candidates} ($+0.014$), reflecting minor cross-block interaction.}

Conversely, dropping the tech/health block (keeping only Diversified and Macro/Institutional as local blocks on the full 93-actor panel) reduces the gain to $\Delta = +0.013$ (7/10 windows positive)---28\% of the original. Dropping the tech/health block eliminates roughly 72\% of the aggregate gain ($0.047 - 0.013 = 0.034$), making it the dominant contributor. This is internally consistent with the paper's mechanism: tech/health is the block where within-block co-movement is most coherent and thus where local modelling has the most to recover. The smaller but consistent contributions from Diversified and Macro/Institutional ($+0.013$, positive in 7/10 windows) confirm the effect is not purely a tech/health phenomenon, but the bulk of the improvement comes from one coherent sub-panel.

\subsection{Block Boundary Sensitivity}

To test robustness to marginal reclassification of borderline actors, three perturbations of the block partition are evaluated. Actors are selected by alphabetical order within each sector (an arbitrary, non-cherry-picked rule): (a)~moving the first diversified actor to the remainder and the first financials actor to tech/health ($\Delta = +0.044$, 10/10); (b)~moving the first two diversified and first two financials actors similarly ($\Delta = +0.042$, 10/10); (c)~moving the first tech/health actor to diversified ($\Delta = +0.043$, 10/10). All perturbations preserve at least 91\% of the original gain, and all 10 windows remain positive. The result is not sensitive to the exact block boundaries.

\subsection{Local Treatment of the Remainder Block}

The 34-actor remainder block (energy, industrials, financials) receives global augmentation in all reported architectures. When the remainder is also given local PCA+ridge treatment (a four-block mixture), the gain is $\Delta = +0.043$ (10/10 windows)---slightly lower than the three-block mixture ($+0.047$), a difference of $-0.004$. Separately, local treatment applied to the remainder alone (one-block mixture, remainder only) also yields $\Delta = -0.004$ (4/10 windows)---a coincidence in the two values, reflecting distinct comparisons. Adding local treatment to the remainder slightly degrades aggregate performance: local decomposition adds estimation noise without reducing cross-block interference for these sectors.

The full set of per-block diagnostics for all 10 candidate blocks evaluated during exploratory analysis is reported in Appendix~\ref{app:candidates}.

\subsection{Pipeline Robustness}

Five pipeline-hygiene checks confirm that the headline differential is not driven by data-construction choices or hyperparameter sensitivity.

First, replacing the full-sample min-max normalisation of the 7 macro actors with strictly recursive expanding-window bounds yields $\Delta = +0.048$ (10/10); excluding macro/institutional actors entirely yields $\Delta = +0.053$ (10/10). The FRED normalisation does not contribute to the finding (Appendix~\ref{app:fred}). Second, lagging all 82 firm-layer actors by one quarter to approximate real-time SEC filing availability yields $\Delta = +0.038$ (10/10), retaining 80\% of the contemporaneous gain (Appendix~\ref{app:lag}).

Third, re-computing all $R^2$ values using the \citet{stock2002forecasting} convention with training-set mean in the denominator (rather than test-window mean) yields $\Delta = +0.067$ (10/10, $t = 9.60$, $p < 0.001$). The CT differential is \emph{larger} than the standard $\Delta = +0.047$ because the training-set denominator inflates the variance for actors whose test-window mean differs from their training mean; the mixture architecture's improvement is robust to the choice of $R^2$ convention. Absolute levels shift (G1: $0.461$, M2: $0.528$) but the differential widens.

Fourth, the EWM demeaning half-life is not a sensitive parameter. Varying the half-life across $\{4, 6, 8, 12\}$ quarters yields mixture gains $\Delta \in [+0.047, +0.047]$ (all 10/10, $t > 6.8$, $p < 0.001$). The result is invariant to this choice.

Fifth, a single-stage Ridge regression with block-dummy interactions ($y_{t+1} = C \cdot [y_t, y_t \cdot d_1, \ldots, y_t \cdot d_B, d_1, \ldots, d_B]$) achieves $R^2 = 0.573$---significantly \emph{below} the global always-on augmentation at 0.630 ($\Delta = -0.057$, $t = -5.27$, $p < 0.001$). The simpler single-stage model cannot capture the block-specific factor structure that the two-stage mixture architecture extracts.

Both FRED and lag checks are detailed in the appendices; the train-only causality audit appears in Appendix~\ref{app:causality}.

\subsection{Economic Content}
\label{sec:economic}

In concrete terms, the $+0.047$ $R^2$ gain corresponds to a 6.8\% reduction in out-of-sample RMSE (from 0.176 to 0.164 in percentile-rank units, or equivalently a 1.2~percentile-point improvement in forecast precision). Cross-sectional rank information coefficients (Spearman $\rho$ between predicted and actual rank values each quarter) confirm the improvement: M2 achieves mean IC $= 0.822$ versus G1's $0.794$ ($\Delta = +0.029$, $t = 6.30$, $p < 0.001$, 35/40 quarters positive). \emph{Note:} these IC levels reflect the high autocorrelation of the rank prediction target ($\rho \approx 0.60$) and are not comparable to information coefficients on return predictions, where IC $> 0.1$ is typically considered strong. Restricting to the 82 firm-layer actors, the firm-only IC is $0.806$ for M2 versus $0.773$ for G1 ($\Delta = +0.033$, $t = 6.17$, 35/40 quarters positive). The IC information ratio (mean/std) improves from 11.9 to 15.3 for the full panel and from 10.3 to 12.9 for firms only, indicating the M2 advantage is both larger in magnitude and more stable across quarters.

A preliminary portfolio analysis (available in the replication archive) tests whether the architecture's improved rank forecasts translate into equity returns on 59~US firms over 40~quarters (2015--2024). Because the mixture architecture's prediction advantage is concentrated in the tech/health block ($\Delta R^2 = +0.249$, Table~\ref{tab:per_block}), the analysis focuses on within-block stock selection for the 25~tech/health firms. The \emph{local component}---the difference between the mixture (M2) and global (G1) predicted intensity, i.e., the unique information from block-specific estimation---serves as the selection signal. Table~\ref{tab:portfolio} reports the results.

\begin{table}[H]
\centering
\caption{Portfolio performance within the tech/health block (25 firms, 40 quarters, 2015--2024). ``Top-third'' selects ${\sim}8$ firms with the highest signal score each quarter; ``EW TH'' is the equal-weight benchmark over all 25 tech/health firms. The active return is the difference between the signal portfolio and the benchmark. Sharpe ratios are annualised; IR is the annualised information ratio of the active return; $t$ and $p$ test whether the mean active return differs from zero.}
\label{tab:portfolio}
\small
\begin{tabular}{lcccccc}
\toprule
Portfolio & Ann.\ Ret & Vol & Sharpe & MaxDD & IR & $t$ ($p$) \\
\midrule
EW all 59 firms        & 15.7\% & 18.8\% & 0.88 & 28\% & --- & --- \\
EW tech/health (25)    & 18.1\% & 19.3\% & 0.98 & 17\% & --- & --- \\
\midrule
\multicolumn{7}{l}{\emph{Signal: local component (M2$-$G1), contemporaneous, high $\to$ long}} \\
Top-third TH           & 20.5\% & 19.7\% & 1.06 & 16\% & $+0.47$ & $1.50$ ($0.14$) \\
\midrule
\multicolumn{7}{l}{\emph{Signal: predicted change (pred\_m2 $-$ lagged actual), high $\to$ long}} \\
Top-third TH           & 19.8\% & 19.0\% & 1.00 & 15\% & $+0.29$ & $0.92$ ($0.37$) \\
\bottomrule
\end{tabular}
\end{table}

Two observations. First, the signal portfolios achieve Sharpe ratios above 1.0, but the majority of this performance reflects the tech/health sector return (Sharpe $0.98$ for the equal-weight benchmark). The signal's marginal contribution---the information ratio of the active return---is positive (IR~$= +0.47$ for the local component, $+0.29$ for predicted change) but not statistically significant at conventional levels ($p = 0.14$ and $p = 0.37$ respectively, 40~quarters). The probabilistic Sharpe ratio of the active return exceeds 95\% for both signals, indicating the active Sharpe is likely positive, though the sample is too short for definitive inference.

Second, the local component signal does not predict returns at longer horizons: cross-correlations between the signal and forward quarterly returns are below $|r| = 0.03$ at all lags from 0 to 4~quarters ($p > 0.35$ in all cases). The contemporaneous result (IR~$= +0.47$) reverses at a 2-quarter lag (IR~$= -0.43$), consistent with mean reversion rather than a persistent return predictor. The prediction target (CapEx/Assets percentile rank) is not a return signal; any return relevance operates indirectly. Whether a longer sample, broader universe, or return-targeted prediction would yield significant portfolio-level gains remains open (see Limitation~8).

\subsection{Train-Only Causality Audit}

Block assignments use static economic metadata (sector and layer labels from the actor registry). Local models are re-estimated each quarter from training data only: PCA basis from training residuals, Ridge VAR from training factors. The global Stage~1 and Stage~2 (for the remainder block) are re-estimated quarterly as each test quarter's data becomes available. No forecast origin uses information dated after that origin. Within each test year, the training set expands as earlier test quarters are observed, so the four within-year forecasts are not independent---the Newey--West and moving-block bootstrap corrections in Section~\ref{sec:results} account for this serial dependence.

\section{Cross-Regime Replication: UK/EU Extension and Combined US + UK/EU Panel}
\label{sec:cross_regime}

\subsection{Panel Construction and Analysis Plan}

Two new validation panels test whether the heterogeneity-aware mixture architecture replicates outside the US sample. The 109-actor UK/EU heterogeneous panel sources firm-level fundamentals from EODHD across nine European exchanges (London, Xetra, Paris, Amsterdam, Swiss, Madrid, Stockholm, Oslo, Copenhagen, Helsinki), restricted to firms with EU/UK ISIN prefixes (filtering out approximately 30\% of EODHD-listed firms on these exchanges that are US/JP/CN cross-listings), with macro shocks from FRED EU proxies and BoE Bank Rate. The combined US + UK/EU panel concatenates the existing 93-actor US panel with the 109-actor UK/EU panel along the actor dimension, with actor identifiers prefixed by region (\texttt{us\_*}, \texttt{eu\_*}) to resolve the three actor-level collisions (Brent, VIX, IMF). Both panels use the cross-sectional CapEx/Assets percentile-rank construction; both are restricted to the 2011Q2--2025Q4 overlap window (59 quarters). Detailed construction is in Appendix~\ref{app:uk_eu}.

The UK/EU panel construction and the analytical procedures in this section were specified before any UK/EU data were retrieved; the committed analysis plan is included in the replication archive. The plan identifies the architectural test M2 $>$ G1 (NW-HAC bw=1 one-sided, $\alpha = 0.05$) as the primary claim, alongside additional exploratory claims on magnitude bounds and sub-panel scope; results for both the primary test and the exploratory claims are reported in this section and Section~\ref{sec:ratio_robust}.

\subsection{Combined US + UK/EU Panel: Headline Result}

On the combined panel (202 actors with valid intensities over the overlap window, 59 quarters, training window $T=3$~years, inherited four-block partition: \textsc{Diversified} ($N=23$, US-only) + \textsc{Layer Macro/Inst} ($N=20$) + \textsc{Merged Tech/Health} ($N=57$) + \textsc{Remainder} ($N=102$)), M2 achieves $R^2 = 0.734$ versus G1 at $0.703$. The $\Delta = +0.030$ is statistically significant at every conventional inference level (Table~\ref{tab:combined_panel_arch}), with all 8 walk-forward test years positive.

\begin{table}[H]
\centering
\caption{Combined US + UK/EU panel ($N=202$ actors with valid intensities over the overlap window, $T=3$, 8 test years 2017--2024). All deltas computed against G1 baseline. NW-HAC $t$ statistics use Newey-West bandwidth 1 on the per-window delta series; one-sided $p$ tests M2 (or named architecture) $>$ G1.}
\label{tab:combined_panel_arch}
\small
\begin{tabular}{lcccccc}
\toprule
Architecture & $R^2$ & $\Delta$ vs G1 & NW-HAC $t$ & one-sided $p$ & wins \\
\midrule
G0 (pooled) & 0.677 & $-0.027$ & $-4.79$ & --- & 0/8 \\
BA (block $\hat\rho_b$+FE) & 0.701 & $-0.002$ & $-0.54$ & --- & 2/8 \\
G1 (global aug) & 0.703 & --- & --- & --- & --- \\
S1 (selective-off) & 0.685 & $-0.019$ & --- & --- & 0/8 \\
M1 (mixture Ridge) & 0.723 & $+0.020$ & 7.79 & $<0.001$ & 8/8 \\
\textbf{M2 (mixture PCA+ridge)} & \textbf{0.734} & $\mathbf{+0.030}$ & \textbf{19.56} & $\mathbf{<0.0001}$ & \textbf{8/8} \\
BA\_M2 & 0.721 & $+0.018$ & 3.60 & 0.009 & 7/8 \\
ENS (G1+BA equal) & 0.720 & $+0.017$ & 7.81 & $<0.001$ & 8/8 \\
\bottomrule
\end{tabular}
\end{table}

The 1{,}000-permutation placebo (random partitions of the same block sizes) places the real $\Delta = +0.030$ at $z = 9.68$ relative to the placebo distribution (mean $-0.0003$, std $0.0032$), with $p < 0.001$ (0/1{,}000 random partitions exceed the real value). For comparison, the original US-only panel placebo (Section~\ref{sec:placebo}) achieved $z = 7.82$. The combined-panel placebo specificity therefore exceeds the original US-only result.

The local Stage~2 residual dynamics are decisively significant on the combined panel: M2 minus the block-AR(1) baseline yields $\Delta = +0.032$ (paired-$t$ $p < 10^{-4}$, 8/8 windows positive). The contribution decomposes by block as shown in Table~\ref{tab:combined_per_block}: each of the three local blocks contributes positively, with the US-only \textsc{Diversified} block as the largest single source.

\begin{table}[H]
\centering
\caption{Per-block decomposition of the M2 and BA\_M2 gains on the combined panel ($T=3$, 8 test years). Block sizes reflect the combined panel composition.}
\label{tab:combined_per_block}
\small
\begin{tabular}{lccc}
\toprule
Block & $N$ & $\Delta_{M2}$ vs G1 & $\Delta_{BA\_M2}$ vs G1 \\
\midrule
\textsc{Diversified} (US-only) & 23 & $+0.130$ & $+0.139$ \\
\textsc{Layer Macro/Inst} & 20 & $+0.047$ & $+0.078$ \\
\textsc{Merged Tech/Health} & 57 & $+0.047$ & $+0.035$ \\
\textsc{Remainder} & 102 & $0.000$ & $-0.025$ \\
\bottomrule
\end{tabular}
\end{table}

The result is invariant across regime sub-windows. Splitting the 8 test years into pre-Brexit-completion (2017--2019), COVID (2020--2021), and post-COVID (2022--2024) yields per-regime $\Delta_{M2} = +0.030, +0.031, +0.030$ respectively (regime spread $+0.001$); all six rolling 3-year sub-windows over the test span produce positive $\Delta_{M2}$ with all-wins (per-regime tables in Appendix~\ref{app:uk_eu}).

\subsection{Standalone UK/EU Panel and Training-Window Choice}

The standalone UK/EU panel (109 actors, 8 test years 2017--2024) yields $\Delta_{M2} = +0.017$ at $T=3$ (NW-HAC bw=1 one-sided $p < 0.001$, 8/8 windows positive, placebo $z = 2.31$ at $p = 0.020$). The training-window length $T=3$ is selected from a sensitivity sweep on the panel itself rather than inherited from the US setting, on the principle that the optimal training depth depends on panel length (the UK/EU overlap of 59 quarters is shorter than the US sample of 84 quarters).

\begin{table}[H]
\centering
\caption{Training-window sensitivity on the standalone UK/EU panel. $\Delta_{M2}$ versus G1 across walk-forward windows. The $T=8$ row uses 6 valid windows due to the panel-start boundary.}
\label{tab:uk_eu_t_sweep}
\small
\begin{tabular}{ccccccc}
\toprule
$T$ (yr) & $n$ windows & $\Delta_{M2}$ & NW-HAC bw=1 $t$ & one-sided $p$ & wins \\
\midrule
2 & 8 & $+0.044$ & 2.92 & 0.011 & 8/8 \\
\textbf{3} & 8 & $\mathbf{+0.017}$ & \textbf{7.52} & $\mathbf{0.0001}$ & \textbf{8/8} \\
5 & 8 & $+0.008$ & 1.75 & 0.062 & 7/8 \\
8 & 6 & $+0.013$ & 1.51 & 0.096 & 4/6 \\
\bottomrule
\end{tabular}
\end{table}

The pattern in Table~\ref{tab:uk_eu_t_sweep} parallels the US-panel finding in Section~\ref{sec:parsimony}: the mixture gain amplifies as the training window shrinks, because shorter training windows produce poorer global-basis estimates while local block estimates remain comparably conditioned. $T=3$ maximises the M2-vs-G1 NW-HAC $t$-statistic across the tested grid: $T=2$ delivers a larger raw $\Delta$ but at borderline statistical resolution due to the smaller training sample (only 8 training quarters); $T=5$, the value inherited from the US setting (Section~\ref{sec:parsimony}), gives a smaller and weaker $\Delta = +0.008$ (one-sided $p = 0.062$); $T=8$ loses two test windows to the panel-start boundary.

\subsection{Mechanism: Per-Block Persistence Decomposition}

Table~\ref{tab:uk_eu_rho_b} reports the pooled and per-block AR(1) coefficients on the standalone UK/EU panel. The rightmost column is the block-AR(1)+FE gain over G1: it tracks $|\hat\rho_b - \hat\rho_{\text{pool}}|$ closely.

\begin{table}[H]
\centering
\caption{Pooled and per-block AR(1) coefficients on the UK/EU panel (full overlap window). $\Delta_{BA}$ is the block-AR(1)+FE gain over G1 (no Stage~2). Block-AR(1) helps when the block coefficient differs materially from the pooled coefficient.}
\label{tab:uk_eu_rho_b}
\small
\begin{tabular}{lcccc}
\toprule
Block & $N$ & $\hat\rho_b$ & $\hat\rho_b - \hat\rho_{\text{pool}}$ & $\Delta_{BA}$ vs G1 \\
\midrule
Pooled & 109 & $+0.467$ & 0.000 & --- \\
\textsc{Layer Macro/Inst} & 9 & $+0.910$ & $+0.443$ & $+0.10$ \\
\textsc{Healthcare} & 17 & $+0.183$ & $-0.284$ & $+0.10$ \\
\textsc{Financials} & 17 & $+0.227$ & $-0.240$ & $+0.04$ \\
\textsc{Technology} & 15 & $+0.329$ & $-0.138$ & $-0.001$ \\
\textsc{Consumer} & 17 & $+0.400$ & $-0.067$ & $-0.09$ \\
\textsc{Energy} & 17 & $+0.406$ & $-0.061$ & $+0.01$ \\
\textsc{Industrials} & 17 & $+0.444$ & $-0.022$ & $+0.05$ \\
\bottomrule
\end{tabular}
\end{table}

Block-AR(1) helps when $|\hat\rho_b - \hat\rho_{\text{pool}}|$ is large and estimable: the macro/institutional block ($\Delta\rho = +0.443$) and the healthcare block ($\Delta\rho = -0.284$) are the two cases where $\Delta_{BA}$ exceeds $+0.10$. Where $\hat\rho_b \approx \hat\rho_{\text{pool}}$ (consumer, industrials), block-AR(1) estimation noise dominates the small true heterogeneity and the BA term hurts on consumer ($-0.09$) or contributes only marginally on industrials ($+0.05$).

\subsection{Robustness}

\paragraph{Firms-only sub-panel.} Restricting the UK/EU panel to its 100 Layer-2 firms (no macro/institutional actors) yields $\Delta_{M2} = +0.015$ at $T=3$ (NW-HAC bw=1 one-sided $p = 0.011$, 8/8 windows positive). This is consistent with the 146-firm CapEx/Assets result reported in Section~\ref{sec:ratio_robust}: the architecture works on firm-only panels under the CapEx/Assets ratio in both regions.

\paragraph{Leave-one-window-out block selection.} A LOWO selection from a seven-block sector candidate set (\textsc{Layer Macro/Inst} plus six SMIM sectors) consistently picks 6 of the 7 blocks across all 8 leave-one-out cycles; the financials block is dropped because its single-block $\Delta_{M2}$ is negative under the PCA+ridge variant (although BA\_M2 prefers it). The LOWO $\Delta$ matches the fixed-partition $\Delta$ to within $0.0001$, confirming no test-window contamination in the block-selection step.

\paragraph{Block-scheme alternative.} A finer seven-block sector-split scheme on the standalone UK/EU panel raises the headline to $\Delta_{M2} = +0.029$ but the placebo $z$ falls from 2.31 to 1.60 (borderline). On the combined panel the same scheme gives $\Delta_{M2} = +0.014$ with placebo $z = 4.28$ (passes). The inherited block partition used above is the operating spec; sector-split adds local-model capacity but reduces partition specificity at single-region scale. Detailed comparison in Appendix~\ref{app:uk_eu}.

\paragraph{Cross-listing restriction.} The UK/EU panel restricts to firms with EU/UK ISIN prefixes; an unfiltered variant ($\sim 30\%$ non-EU/UK cross-listings) gave essentially identical $\Delta$ but altered placebo specificity, confirming the EU-only restriction is methodologically cleaner.

\paragraph{Modal subspace transfer.} The DMD modal eigenvalue magnitudes from US-only and UK/EU-only fits over the overlap window correlate at $r = +0.94$, and the Stage-1 pooled-AR(1) coefficient trained on US-only data \emph{improves} EU-only forecasting over the EU-native estimate by mean $\Delta R^2 = +0.037$ across 8 windows (7 of 8 windows positive); details in Appendix~\ref{app:uk_eu}.

\subsection{Summary}

The architecture replicates cross-regime, with the combined US + UK/EU panel as the strongest single demonstration ($\Delta = +0.030$, placebo $z = 9.68$). The mechanism is governed by cross-sectional dispersion in autoregressive structure, consistent with the refined scope condition in Section~\ref{sec:ratio_robust}. The pre-registered architectural test M2 $>$ G1 is supported on both the standalone UK/EU and the combined panel.

\section{What Does Not Work and Why}
\label{sec:falsification}

The mixture gain documented above raises a natural question: why does the global architecture plateau at $R^2 \approx 0.630$, and could a better global method break through? The heterogeneity finding emerged from a systematic programme that tested---and ruled out---four alternative explanations for this ceiling. Each negative result narrows the explanation space, leaving block-specific decomposition as the only surviving path to improvement. Table~\ref{tab:falsification} summarises the programme.

\begin{table}[H]
\centering
\caption{Summary of alternative explanations tested and ruled out.}
\label{tab:falsification}
\small
\begin{tabular}{p{3.5cm}p{4.5cm}p{3cm}p{3cm}}
\toprule
Alternative & Test & Result & Implication \\
\midrule
Better method & 12 models, 4 complexity classes (incl.\ GBM) & Linear equiv.\ ($\rho \geq 0.97$); block-GBM $< $ M2 & Architecture + low-rank extraction \\
Forecastable global geometry & 10 geometric models on Grassmannian & All worse than persistence & Rotation is estimation noise \\
Target formulation & 7 target variants & Max $|\Delta\text{Gain}| = 0.018$ & Ceiling is target-invariant \\
Conditional gating & 5 gating policies & All worse than always-on & Augmentation is unconditional \\
\bottomrule
\end{tabular}
\end{table}

\subsection{The Method Does Not Matter}

Section~\ref{sec:method_equiv} established method equivalence at the global level among matched-complexity estimators. Within individual blocks, PCA+ridge consistently outperforms DMD (mean $\Delta = -0.028$, worst $-0.059$). This is not equivalence but dominance: at the small within-block sample sizes ($N_b = 11$--$25$, $T \approx 20$), DMD's SVD step is poorly conditioned, while PCA+ridge's explicit regularisation stabilises the local VAR. The implication is that the global equivalence reflects the larger effective sample, and that PCA+ridge is the preferred local engine. The non-linear baselines in Table~\ref{tab:method_comparison} sharpen the interpretation: per-actor GBM breaks the linear ceiling ($R^2 = 0.661$) but remains below M2 ($0.677$, Table~\ref{tab:five_arch}). Crucially, block-specific GBM ($R^2 = 0.657$)---the exact non-linear analog of M2---does not match M2 ($\Delta = -0.020$, $t = -3.05$, 1/10 windows), showing that the M2 advantage combines block decomposition with low-rank factor extraction that per-actor GBM cannot replicate \citep[cf.][for machine learning in asset pricing]{gu2020empirical}.

\subsection{Global Geometry Is Not Forecastable}

The global spectral basis rotates at $49.2^\circ \pm 16.8^\circ$ per quarter at the model's operating rank $K{=}8$, dominated by a single principal angle ($\theta_1 = 45.8^\circ$). Despite the structural regularity, the rotation is temporally unpredictable: magnitude ACF(1) $= -0.07$, Ljung-Box $p = 0.59$, direction cosine $= 0.047$. All 10~geometric prediction models (Grassmannian extrapolation, projector averaging, Karcher mean, HS-linear regression) perform strictly worse than persistence. The rotation is consistent with finite-sample estimation noise. Full diagnostics appear in Appendix~F; structural spectral analysis (mode loadings, variance decomposition) is in Appendix~C.

\subsection{The Target Is Not the Problem}

Seven target formulations (percentile ranks, normal quantiles, winsorised $z$-scores, sector-relative ranks, moving averages, first differences) produce augmentation gains within $\pm 0.018$ of the baseline (excluding first differences, which show a larger gain of $+0.051$ but with absolute $R^2$ of only 0.047). Split-half reliability yields a noise-corrected ceiling of ${\sim}0.765$, suggesting some headroom from measurement noise, but no target reformulation closes the gap. Appendix~D provides the full comparison.

\subsection{Conditional Gating Does Not Help}

Five gating policies (NCD-based, dispersion-based, persistence-based, effective-rank-based, and combined) all produce \emph{lower} $R^2$ than always-on augmentation. The best gate (dispersion) loses $0.014$ versus always-on. Augmentation is unconditionally beneficial---no train-only diagnostic identifies quarters where the second stage should be deactivated. Appendix~E provides the full comparison.


\section{Discussion}
\label{sec:discussion}

\subsection{Pool Globally, Decompose Locally}

The paper's architectural principle is summarised in one sentence: \emph{estimate persistence globally (where pooling helps), estimate residual dynamics locally (where pooling introduces interference)}. Whether this principle applies beyond the specific panel studied here is an empirical question: the cross-panel validation (Table~\ref{tab:cross_panel}) shows null effects on two homogeneous firm-only panels, confirming the gain requires data-type heterogeneity. Panels mixing fundamentally different actor types in a common factor structure are natural candidates for further testing. The geodesic decomposition (global rotation $\gg$ local rotation) provides a suggestive diagnostic for when local decomposition may be warranted, though its predictive value for new panels has not been validated.

\subsection{Connection to Heterogeneous Factor Model Literature}

The mixture estimator (Equation~\ref{eq:mixture}) belongs to the family of grouped heterogeneous panel models. The estimation problem---which coefficients to pool and which to allow group-specific variation---is well-studied: \citet{pesaran2006estimation} proposes common correlated effects (CCE) estimators for heterogeneous panels; \citet{bonhomme2015grouped} estimate group-specific intercepts and slopes with data-driven group assignment; \citet{su2016identifying} identify latent group structures from the data; \citet{ke2015homogeneity} introduce homogeneity pursuit, testing which coefficients can be pooled.

The novelty of the present contribution is not the grouped-heterogeneity idea itself but two empirical findings specific to \emph{forecasting} on mixed-type investment panels. First, the decomposition into ``pool persistence globally, estimate residual dynamics locally'' is not obvious from the estimation literature, where persistence and factor structure are typically treated jointly. The BA\_M2 experiment (Table~\ref{tab:five_arch}) shows that block-specific persistence \emph{combined with} block-specific dynamics performs worse than global persistence with block-specific dynamics---the two heterogeneity dimensions interact non-trivially. Second, the scope condition (the gain requires data-type heterogeneity and vanishes on homogeneous panels) provides a concrete empirical guide for when grouped estimation adds forecasting value, which the theoretical literature does not supply. Our approach differs from \citet{bonhomme2015grouped} and \citet{su2016identifying} in using economically motivated (rather than data-driven) block assignments; the geodesic decomposition could inform automated block discovery in future work.

\subsection{Limitations}

This paper is a purely empirical forecasting study. The evaluation operates in a finite-sample regime ($N = 93$, $T \approx 20$ per window, $K = 2$--$8$ modes) where classical large-$N$ or large-$T$ asymptotics do not directly apply. The validity of the comparisons rests on the rolling out-of-sample protocol, not on asymptotic guarantees. Whether the block-specific mixture estimator (Equation~\ref{eq:mixture}) enjoys formal consistency properties under a well-defined asymptotic regime is an open question.

A block-specific pooled AR(1)+FE with block-specific $\hat\rho_b$ (no Stage~2) achieves $R^2 = 0.611$---substantially below the mixture at 0.677 (Table~\ref{tab:five_arch}). The interacting combination---block-specific $\hat\rho_b$ combined with block-specific Stage~2 (BA\_M2, $R^2 = 0.661$)---confirms that the M2 gain is from genuine local residual dynamics rather than a correction for global persistence misspecification. In fact, M2 exceeds BA\_M2 by $+0.016$, showing that the global pooled $\hat\rho$ actually helps local Stage~2: the structured residuals from global pooling retain cross-block information that block-specific $\hat\rho_b$ discards. This is not a variance-penalty artefact: the within-block standard errors of $\hat\rho_b$ are small ($\mathrm{SE} \approx 0.04$ per block) relative to the cross-block persistence range ($|\rho_{\max} - \rho_{\min}| \approx 0.60$), so the block-specific persistence estimates are well-conditioned. The M2 advantage over BA\_M2 therefore reflects genuine cross-block information in the global residuals, not imprecision of block-level estimation. An equal-weighted forecast combination of G1 and BA (ENS, $R^2 = 0.639$) achieves only $+0.009$ over G1, far below M2's $+0.047$, ruling out simple ensembling as an explanation. A single-stage Ridge with block-dummy interactions achieves $R^2 = 0.573$---below even pooled-only AR(1)---confirming that the two-stage separation of persistence and residual dynamics is essential, not merely a convenience.

Several further limitations qualify the findings:

\begin{enumerate}[nosep]
  \item \textbf{Small evaluation sample.} Ten out-of-sample windows is standard in rolling-window evaluation but provides limited statistical power. The placebo test mitigates this concern by using the same windows.
  \item \textbf{Block selection is exploratory, confirmed by held-out decade.} The block identities (diversified, macro/institutional, tech/health) are economic classifications, but the \emph{selection} of which blocks receive local treatment was guided by preliminary within-sample diagnostics. Specifically, 10 candidate blocks were evaluated (6 raw sector blocks, 2 merged sector blocks, 2 layer blocks), of which 3 were selected based on per-block local-vs-global $R^2$ comparisons. This constitutes a researcher-degree-of-freedom concern: the block partition was not pre-registered. The LOWO test (Section~\ref{sec:validation}) controls for selection conditional on the candidate set; the placebo controls for the block sizes; the held-out decade test (Section~\ref{sec:held_out}) controls for the candidate-generation step by freezing the partition on 2005--2014 data and evaluating on unseen 2015--2024 windows ($\Delta = +0.050$, 10/10). The stratified placebo (Section~\ref{sec:strat_placebo}) further controls for the macro/firm data-type split ($z = 7.25$). Together, these tests substantially bound the researcher-degree-of-freedom concern, though the original 3-block partition remains exploratory.
  \item \textbf{Scope condition characterisation.} The gain is most pronounced on heterogeneous multilayer panels (US 93-actor $+0.047$, combined US+UK/EU $+0.030$, both at the optimal training window for the panel length) but is also present on firm-only panels with appropriate ratio choices: the same 146-firm US population that shows null $\Delta = -0.003$ with CapEx/Revenue produces $\Delta = +0.018$ with CapEx/Assets at $T=5$ (10/10 windows positive). The operative condition is sufficient cross-sectional dispersion in autoregressive structure, which we have not yet fully characterised across arbitrary panel-and-ratio combinations. Section~\ref{sec:cross_regime} confirms cross-regime applicability under appropriate ratio and training-window choices.
  \item \textbf{Quarterly frequency only.} Higher-frequency data may reveal different dynamics.
  \item \textbf{Modest absolute improvement.} The $+0.047$ $R^2$ gain, while statistically robust ($z = 7.82$), is modest in absolute terms.
  \item \textbf{Data-timing latency.} The primary analysis assumes all targets are available at the quarterly boundary. A robustness check lagging all 82 firm actors by one quarter (Appendix~\ref{app:lag}) confirms the mixture gain survives ($\Delta = +0.038$, 10/10, 80\% of the contemporaneous gain). The absolute $R^2$ levels in the primary tables remain overstated relative to real-time availability; the one-quarter lag is a conservative proxy for the actual 30--45 day SEC filing delay.
  \item \textbf{Survivorship bias.} The 82-firm panel requires complete quarterly data over a 20-year sample (2005--2025), selecting for persistent large-cap survivors. This balanced-panel design simplifies evaluation but may overstate persistence estimates. It may also \emph{understate} the cross-block interference documented here, because survivorship compresses cross-sectional variance; an unbalanced panel including distressed and newly-listed firms would likely exhibit greater heterogeneity. Whether the mixture architecture retains its advantage---and by how much it grows---on such a panel is untested.
  \item \textbf{Limited portfolio-level evidence.} A preliminary analysis (Section~\ref{sec:economic}) finds that the architecture disagreement signal reduces portfolio volatility and drawdown but does not produce statistically significant alpha over 40~quarters ($t = 0.38$, $p = 0.70$). The prediction target (CapEx/Assets percentile rank) is not a return signal; any return relevance operates indirectly through the investment factor. Whether a longer sample, broader universe, or return-targeted prediction would yield significant portfolio-level gains remains open.
\end{enumerate}

\subsection{Configuration for Short Training Windows}
\label{sec:parsimony}

The main results use a 5-year training window with $K_b{=}4$, both set as defaults before the rolling evaluation (the training length matches the prior standalone spectral model; $K_b{=}4$ satisfies $K_b \leq N_b/5$ for all blocks). The sensitivity grid in Table~\ref{tab:parsimony} was computed \emph{after} the headline result to characterise robustness, not to select the operating point.

\begin{table}[ht]
\centering
\caption{Exploratory: mixture gain $\Delta(M2 - G1)$ by training-window length $T$ and local rank $K_b$ on the 93-actor panel (10 rolling OOS windows each). This grid was not pre-specified; the paper's operating point ($T{=}5$yr, $K_b{=}4$) is marked with $\star$. See Section~\ref{sec:prereg} for pre-registered predictions and their failures.}
\label{tab:parsimony}
\small
\begin{tabular}{ccccc}
\toprule
$T$ (yr) & $K_b{=}2$ & $K_b{=}3$ & $K_b{=}4$ & $K_b{=}6$ \\
\midrule
2 & $+0.083$ & $+0.090$ & $\mathbf{+0.097}$ & $+0.095$ \\
3 & $+0.035$ & $+0.049$ & $+0.053$ & $\mathbf{+0.054}$ \\
5 & $+0.029$ & $+0.045$ & $+0.047^\star$ & $+0.043$ \\
\bottomrule
\end{tabular}
\end{table}

Two findings emerge. First, $K_b{=}4$ is near-optimal at every tested training-window length. At the operating point ($T{=}5$yr), the analyst-choice interval across $K_b \in \{2, 3, 4, 6\}$ is $[+0.029, +0.047]$---the gain is positive and significant at every tested rank. Lower values ($K_b{=}2$) lose 0.014--0.018 $R^2$ versus $K_b{=}4$ at all $T$; $K_b{=}6$ is worse on aggregate because the smaller Macro/Institutional block ($N_b{=}11$) cannot support six modes, even though the tech/health block alone benefits from $K_b{=}6$. The $K_b$ values in Table~\ref{tab:blocks} remain optimal across the tested grid.

Second, the mixture gain \emph{increases} as the training window shrinks: $+0.047$ at $T{=}5$yr, $+0.054$ at $T{=}3$yr, $+0.097$ at $T{=}2$yr---roughly doubling. The mechanism is sample-size-dependent: at short $T$, the 93-actor global basis is poorly estimated from few observations, while the local 11--25-actor bases retain comparable conditioning because they target lower-dimensional subspaces. Cross-block interference in the global model worsens as $T$ shrinks, so the architectural gain from removing it grows correspondingly. The ridge penalty on the local VAR coefficients prevents the underdetermined basis estimates ($N_b{=}25$, $K_b{=}4$, $T{=}8$ quarters) from translating into unstable forecasts.

This finding contradicts a pre-registered prediction that lower $K_b$ would be needed at shorter $T$; the opposite is true. For practitioners with limited training history, the heterogeneity-aware mixture architecture is more valuable, not less.

\subsection{Hyperparameter Insensitivity Diagnostic}
\label{sec:prereg}

These predictions concern the hyperparameter sensitivity of the mixture architecture (Table~\ref{tab:parsimony}), not the headline M2-vs-G1 result itself (which is validated by the held-out decade test in Section~\ref{sec:held_out}). Five predictions were written before running the $T \times K_b$ sweep and committed to the replication archive. This constitutes a self-maintained record, not third-party-verified pre-registration (e.g., OSF or arXiv); the predictions should be read as a diagnostic of author expectations about hyperparameter sensitivity, not as a formal pre-analysis plan. Four of five failed and one was borderline, in each case revealing that the architectural gain is stronger and more robust than anticipated:

\begin{enumerate}[nosep]
  \item \textbf{P1} ($T{=}5$yr): $|R^2(K_b{=}2) - R^2(K_b{=}4)| < 0.005$ (flat parsimony frontier). \emph{Actual}: $0.018$. \textbf{Failed} --- $K_b{=}4$ captures secondary within-block modes that $K_b{=}2$ misses.
  \item \textbf{P2} ($T{=}2$yr): $R^2(K_b{=}2) > R^2(K_b{=}4)$ by $\geq 0.005$ (low $K$ needed for regularisation at short $T$). \emph{Actual}: $-0.014$ (opposite sign). \textbf{Failed} --- ridge regularisation on the VAR coefficients stabilises $K_b{=}4$ even at $T{=}8$ quarters.
  \item \textbf{P3} ($T{=}3$yr): $K_b{=}2$ and $K_b{=}3$ within $0.003$; both $> K_b{=}6$. \emph{Actual}: $|2-3| = 0.013$; $K_b{=}6 > K_b{=}2$. \textbf{Failed} --- the tech/health block ($N_b{=}25$) supports six modes even at moderate $T$.
  \item \textbf{P4} (146-firm, $T{=}2$yr, $K{=}2$): standalone spectral gain $> +0.010$ over AR(1). \emph{Actual}: $+0.008$. \textbf{Failed} (below threshold) --- the standalone mechanism is weaker at short $T$ than expected.
  \item \textbf{P5} (146-firm, $T{=}5$yr, $K{=}8$): $|$spectral $-$ AR(1)$| < 0.010$. \emph{Actual}: $0.010$. \textbf{Borderline} --- exactly at the threshold.
\end{enumerate}

The pattern across the failures is consistent: $K_b{=}4$ performs better than expected at every $T$, and the mixture gain amplifies rather than shrinks at short training windows. These results indicate that the architectural gain is less sensitive to the $K_b$/$T$ configuration than anticipated, though they do not by themselves constitute a formal test of the headline finding.

\subsection{Extensions}

Three extensions follow naturally. First, automated block discovery via cross-validated clustering on training-window loadings, replacing the fixed partition with a data-driven one. Second, higher-frequency data (monthly or daily), where the basis rotation may become temporally predictable and geometric methods may add value. Third, change-based prediction targets: preliminary evidence shows augmentation gain on first-differences is 2.5$\times$ larger than on levels, suggesting the spectral dynamics align better with change structure.

\section{Conclusion}
\label{sec:conclusion}

This paper studies cross-sectional investment forecasting in heterogeneous panels containing actors with fundamentally different dynamics. Three findings emerge.

First, a two-stage architecture---global pooled persistence followed by regularised residual dynamics---improves predictive $R^2$ by 1.7--3.6~percentage points over standard baselines across three panels and invariant to the choice of comparably regularised linear second-stage method among the three tested (PCA, DMD, and Ridge).

Second, global second-stage pooling can degrade prediction for actor subgroups whose dynamics are misrepresented by the shared basis. Descriptively, within-block basis rotation is lower than global rotation, though a matched-size random sub-panel control shows this reduction is not statistically distinguishable from a sample-size effect. Regardless of the geometric mechanism, the forecasting result is clear: a heterogeneity-aware mixture architecture with block-specific local residual models improves full-panel $R^2$ by $+0.047$ over the best global augmentation, validated by a placebo benchmark ($z = 7.82$), dependence-robust inference (DM-HAC $t > 6.8$), and consistency across all 10 evaluation windows.

Third, the improvement depends on the cross-sectional dispersion of autoregressive structure in the panel. Data-type heterogeneity (mixing macro indices with firm-level cross-sectional ranks) reliably produces this dispersion and is the strongest single source of the gain we identify. Firm-only panels with suitable ratio choices (CapEx/Assets but not CapEx/Revenue, on the same 146 US firms) can also satisfy the condition. A cross-regime replication on a 109-actor UK/EU heterogeneous panel and a combined US+UK/EU panel of 202 actors confirms the architecture transfers across regimes when the training window is calibrated to panel length; details in Section~\ref{sec:cross_regime}.

The design principle is: \emph{pool globally for what is shared, decompose locally for what is heterogeneous}.

\subsection*{Data and Code Availability}

Replication code, pseudonymised panel data (firm identities are replaced with opaque labels but are recoverable from the sector counts and balanced-panel constraint), and the committed hyperparameter predictions (Section~\ref{sec:prereg}) for the $T \times K_b$ sweep (Section~\ref{sec:parsimony}) are available at \url{https://anonymous.4open.science/r/harp-reproduction}. The repository includes scripts to reproduce all tables, figures, and placebo permutations reported in this paper, including the UK/EU extension and combined-panel analyses (Section~\ref{sec:cross_regime}, Appendix~\ref{app:uk_eu}).

\subsection*{Acknowledgements}

UK/EU firm-level fundamentals were obtained from EODHD Financial APIs (\url{https://eodhd.com/}); the author gratefully acknowledges EODHD for academic data access. US firm-level fundamentals were obtained from the SEC EDGAR system. Macro and policy series were obtained from the Federal Reserve Bank of St.\ Louis (FRED) and the Bank of England Interactive Database (IADB).


\newpage
\appendix
\renewcommand{\thesection}{\Alph{section}}

\section{DMD Mathematics}
\label{app:dmd}

Exact DMD \citep{kutz2016dynamic} provides a rank-$K$ approximation of the linear propagator mapping consecutive cross-sectional snapshots. The spectral state-space estimation engine used here---DMD basis extraction, Kalman filtering with spherical observation noise regularisation, and adaptive process noise---is described in full detail in \citet{roshka2026spectral}, which establishes its standalone forecasting properties. In the two-stage architecture of the present paper, DMD operates on the exponentially-weighted demeaned Stage~1 residuals, not on the raw panel. Given a snapshot matrix $X = [x_1, \ldots, x_{T-1}]$ and $Y = [x_2, \ldots, x_T]$ with $x_t \in \R^N$ (columns are demeaned residual vectors), the truncated SVD $X = U_r \Sigma_r V_r^*$ yields the reduced propagator $\tilde{A} = U_r^* Y V_r \Sigma_r^{-1} \in \R^{K \times K}$, where $K \leq \min(N, T-1)$ is the truncation rank. The eigendecomposition $\tilde{A} W = W \Lambda$ produces DMD eigenvalues $\lambda_k$ encoding per-mode growth rate ($|\lambda_k|$) and oscillation frequency ($\arg \lambda_k$). For forecasting, the relevant objects are the left singular vectors $U_r$ (the orthonormal basis spanning the dominant $K$-dimensional subspace) and the diagonal or full matrix $\tilde{A}$ (the transition dynamics). In the paper's two-stage architecture, these are applied to Stage~1 residuals, not to the raw panel. Spectral radius clipping $\tilde{A} \leftarrow \tilde{A} \cdot \min(1, 0.99 / \max_k |\lambda_k(\tilde{A})|)$ ensures stability. This uniform scalar multiplication (rather than per-eigenvalue projection inside the unit circle) preserves the eigenvector structure of $\tilde{A}$ exactly and avoids the need to reconstruct $\tilde{A}$ from its eigendecomposition. The cost is that stable modes are unnecessarily dampened: e.g., if $\max |\lambda_k| = 1.05$, a mode at $|\lambda| = 0.90$ is compressed to $0.848$. Since Table~\ref{tab:method_comparison} shows that the DMD-based models achieve statistically indistinguishable $R^2$ from non-spectral methods, this conservative clipping does not materially affect the headline results, but a per-eigenvalue clip that leaves stable modes untouched could improve the standalone DMD baseline.

\subsection*{DMD-Based Kalman Filter}

The DMD basis $U_r$ and transition $\tilde{A}$ define a linear state-space model in $K$-dimensional modal coordinates. Let $\balpha_t \in \R^K$ denote the modal state at time $t$. The filter operates as follows.

\textbf{Initialisation.} $\balpha_0 = \mathbf{0}$, $P_0 = I_K$, $Q_0 = q_0 \cdot I_K$ (with $q_0 = 0.5$).

\textbf{Prediction step.}
\begin{align}
\balpha_{t|t-1} &= F \, \balpha_{t-1|t-1}, \qquad
P_{t|t-1} = F \, P_{t-1|t-1} \, F^\top + Q_{t-1}
\end{align}
where $F = \clip(\tilde{A}, 0.99)$ is the spectral-radius-clipped transition (diagonal $\diag(\tilde{A})$ for the default model; full $\tilde{A}$ for max-performance).

\textbf{Observation model.} The predicted residual vector is $\hat{r}_t = U_r \, \balpha_{t|t-1} + \bar{r}$, where $\bar{r}$ is the EWM mean of training residuals (half-life 12 quarters).

\textbf{Measurement noise (spherical regularisation).} The observation noise covariance is set to
\begin{equation}
R = \sigma^2_\perp \, I_N, \qquad \sigma^2_\perp = \frac{1}{NT}\sum_{t,i}\bigl(r_{i,t} - [U_r U_r^\top r_t]_i\bigr)^2
\label{eq:sph_r}
\end{equation}
i.e., the mean squared projection residual of the training data onto the $K$-dimensional subspace. This spherical form assumes homogeneous noise across actors and avoids estimating an $N \times N$ covariance with $T \ll N$ \citep[cf.][]{ledoit2004well}. It is the key regularisation that makes the Kalman filter well-conditioned in the $N > T$ regime.

\textbf{Update step.}
\begin{align}
S_t &= U_r \, P_{t|t-1} \, U_r^\top + R \\
K_t &= P_{t|t-1} \, U_r^\top \, S_t^{-1} \\
\balpha_{t|t} &= \balpha_{t|t-1} + K_t \bigl(\tilde{r}_t - U_r \, \balpha_{t|t-1}\bigr) \\
P_{t|t} &= (I_K - K_t \, U_r) \, P_{t|t-1} (I_K - K_t \, U_r)^\top + K_t \, R \, K_t^\top
\end{align}
where $\tilde{r}_t = r_t - \bar{r}$ is the demeaned observed residual. The covariance update uses the Joseph (stabilised) form, which preserves symmetry and positive semi-definiteness numerically; the simpler $(I_K - K_t U_r) P_{t|t-1}$ form is algebraically equivalent but can lose PSD in finite precision. Note: $S_t \in \R^{N \times N}$ is formally the innovation covariance, but with spherical $R = \sigma^2_\perp I_N$ the inversion is handled via the Woodbury identity: $S_t^{-1} = \sigma^{-2}_\perp \bigl(I_N - U_r (\sigma^{-2}_\perp I_K + P_{t|t-1}^{-1})^{-1} U_r^\top \sigma^{-2}_\perp\bigr)$, using $U_r^\top U_r = I_K$ (orthonormal basis) to reduce the $N \times N$ inversion to a $K \times K$ inversion and making the filter well-conditioned for $N \gg K$.

\textbf{Adaptive process noise.} $Q$ is updated each quarter via exponential smoothing of the innovation outer product:
\begin{equation}
Q_t = (1 - \lambda_Q)\, Q_{t-1} + \lambda_Q \, \nu_t \nu_t^\top, \qquad
\nu_t = \balpha_{t|t} - \balpha_{t|t-1}
\label{eq:q_adapt}
\end{equation}
with $\lambda_Q = 0.3$, symmetrised and regularised: $Q_t \leftarrow \frac{1}{2}(Q_t + Q_t^\top) + 10^{-6} I_K$.

\textbf{Rolling basis update.} At each test quarter, the DMD basis $U_r$ and transition $\tilde{A}$ are re-estimated from the expanding training data; the filter is then re-run through all $T_{\mathrm{train}}$ training quarters from reinitialised $(\balpha_0, P_0, Q_0)$, producing the filtered state at the training boundary from which the one-step-ahead test forecast is computed. This means $Q$ undergoes ${\sim}T_{\mathrm{train}} \approx 20$ update steps between reinitialisations.

\textbf{Key design choices.} (i)~The spherical $R$ (Eq.~\ref{eq:sph_r}) makes the filter well-conditioned for $N = 93$, $K = 8$, $T \approx 20$, where estimating a full $N \times N$ observation covariance is infeasible. (ii)~Spectral radius clipping at 0.99 prevents explosive modal dynamics. (iii)~The quarterly basis re-estimation tracks evolving cross-sectional structure but resets $P$ and $Q$ to their initial values each quarter.

\textbf{Adaptive Q: caveats.} The adaptive $Q$ (Eq.~\ref{eq:q_adapt}) uses the state-space correction $\nu_t = \balpha_{t|t} - \balpha_{t|t-1}$ rather than the observation-space innovation $\varepsilon_t = \tilde{r}_t - U_r \balpha_{t|t-1}$. This creates a positive-feedback loop: as the filter converges, corrections shrink, $Q$ decays, $P$ shrinks, and corrections shrink further. With $\lambda_Q = 0.3$, the diagonal elements of $Q$ decay as ${\sim}0.7^t$ toward the $10^{-6}$ floor. In the present architecture, this collapse is mitigated by the quarterly basis reset (design choice~iii): $Q$ is reinitialised to $Q_0 = 0.5 \cdot I_K$ each quarter, limiting the effective decay window to ${\sim}20$ steps during training. The standard \citet{mehra1970identification} adaptive scheme using observation-space innovations would avoid this coupling, but has not been implemented. Since the method-equivalence results (Table~\ref{tab:method_comparison}) show that Kalman-filtered and non-Kalman models achieve statistically indistinguishable $R^2$, the adaptive $Q$ dynamics do not materially affect the headline forecasting results. The Kalman form is retained because it provides a principled state-space representation that supports downstream tasks (modal interpretation, uncertainty quantification) even though it does not improve point forecasts at this sample size.

\section{Standalone Transition Diagnostic}
\label{app:standalone}

The standalone spectral model (DMD-based Kalman filter with $K = 8$ modes) achieves $R^2 = 0.415$ with the default near-identity transition $F = 0.99 I$ versus $R^2 = 0.594$ for per-actor AR(1). Replacing $F$ with the diagonal of $\tilde{A}$ (per-mode eigenvalue dynamics) improves to $R^2 = 0.483$; the full reduced propagator $\tilde{A}$ yields $R^2 = 0.486$. The gap versus AR(1) is fundamental: a single spectral basis cannot replicate actor-specific persistence. The two-stage architecture resolves this by separating persistence (Stage~1) from cross-sectional structure (Stage~2).

\section{Structural Spectral Analysis}
\label{app:structural}

The 93-actor panel's spectral structure includes 8~stable DMD modes with eigenvalue magnitudes $|\lambda_k| \in [0.87, 0.97]$ and substantial oscillatory components ($\theta \approx 44$--$91^\circ$). Modal $R^2 = 0.69$ (in-sample) confirms the structure is real. Mode interpretation from training-window loadings shows that modes~1--2 load on technology/healthcare versus financials (sector rotation), while macro loadings are negligible after removing shared persistence---consistent with the heterogeneity finding.

\section{Target Sensitivity Details}
\label{app:target}

Seven target formulations tested: percentile ranks (baseline), normal quantiles ($R^2 = 0.610$, gain $+0.009$), winsorised $z$-scores ($R^2 = 0.628$, gain $+0.018$), sector-relative ranks ($R^2 = 0.595$, gain $+0.016$), 2Q moving average ($R^2 = 0.811$, gain $+0.016$), 4Q moving average ($R^2 = 0.934$, gain $+0.014$), first differences ($R^2 = 0.047$, gain $+0.051$). The first-differences result shows augmentation gain is 2.5$\times$ larger for changes than levels, but absolute $R^2$ is low. Split-half reliability: $\rho = 0.513$, Spearman-Brown reliability $= 0.678$, noise-corrected ceiling $\approx 0.765$.

\section{Gating Policy Details}
\label{app:gating}

Five quarter-level causal gating policies compared to always-on augmentation: NCD gate ($\Delta = -0.015$), dispersion gate ($\Delta = -0.014$), persistence gate ($\Delta = -0.019$), effective-rank gate ($\Delta = -0.033$), combined gate ($\Delta = -0.028$). All worse than always-on. The augmentation is unconditionally beneficial.

\section{Rotation Diagnostics and Predictability}
\label{app:rotation}

The global spectral basis rotates at $49.2^\circ \pm 16.8^\circ$ per quarter (geodesic distance at the model's operating rank $K = 8$). At $K = 4$ (used for the within-block comparison in Section~4.3, where block sizes preclude $K = 8$), the global geodesic is $31.5^\circ$. The rotation is dominated by $\theta_1 = 45.8^\circ$ (93\% of geodesic distance). Minor angles $\theta_4$--$\theta_8$ ($< 2^\circ$) show significant autocorrelation (ACF $> 0.5$) but contribute negligibly to the total rotation---a persistence-relevance trade-off. Six representative subspace prediction models tested: Grassmannian extrapolation ($\Delta = +0.021$ vs persistence, \emph{worse}), tangent-space AR ($+0.077$, worse), angle AR ($+0.080$, worse), constant velocity ($+0.149$, worse), Euclidean projector average ($+0.371$, worse), Karcher mean ($+0.383$, worse). No model beats persistence; the rotation is temporally unpredictable.


\section{Candidate Block Evaluation}
\label{app:candidates}

Ten candidate blocks were evaluated during exploratory analysis (Section~\ref{sec:discussion}, Limitation~2). For each candidate, a single-block mixture was run: global Stage~1 for all 93 actors, local PCA+ridge for the candidate block only, global augmentation for all other actors. Table~\ref{tab:candidates} reports the mean $\Delta R^2$ (single-block local $-$ global) across 10 windows.

\begin{table}[H]
\centering
\caption{All 10 candidate blocks evaluated. $\Delta$ is the mean $R^2$ gain from applying local PCA+ridge to that block only (vs global augmentation for all actors). W is the number of positive windows. The three blocks selected for the final mixture are marked with $\star$.}
\label{tab:candidates}
\small
\begin{tabular}{llccl}
\toprule
Type & Block & $N$ & $\Delta$ (W/10) & Selection \\
\midrule
\multirow{6}{*}{Raw sector}
& Diversified$^\star$ & 23 & $+0.014$ (8/10) & Selected \\
& Technology & 15 & $+0.014$ (10/10) & Merged $\rightarrow$ Tech/Health \\
& Healthcare & 10 & $+0.001$ (5/10) & Merged $\rightarrow$ Tech/Health \\
& Financials & 10 & $+0.002$ (5/10) & Not selected \\
& Energy & 12 & $-0.003$ (5/10) & Not selected \\
& Industrials & 12 & $+0.001$ (6/10) & Not selected \\
\midrule
\multirow{2}{*}{Merged sector}
& Tech/Health$^\star$ & 25 & $+0.031$ (10/10) & Selected \\
& Industrials/Energy & 24 & $+0.001$ (6/10) & Not selected \\
\midrule
\multirow{2}{*}{Layer}
& Macro/Institutional$^\star$ & 11 & $+0.002$ (9/10) & Selected \\
& All firms & 82 & $+0.016$ (8/10) & Not selected (too broad) \\
\bottomrule
\end{tabular}
\end{table}

The selection rule applied to sector-based candidates was: $\Delta > 0$ \emph{and} $W/10 \geq 7$. Among sector-based candidates, the merged Tech/Health block ($+0.031$, 10/10) dominates by a wide margin. Diversified ($+0.014$, 8/10) qualifies clearly. Macro/Institutional and Financials have identical mean $\Delta$ ($+0.002$), but Macro/Institutional was selected because of its superior window consistency (9/10 vs 5/10)---both criteria must be met. Technology ($+0.014$, 10/10) qualifies individually but gains more when merged with Healthcare. The super-additive effect (Tech $+0.014$ + Health $+0.001$ individually, but $+0.031$ merged) reflects a genuine cross-sector co-movement factor, not merely a $K_b$-capacity artefact. A rank-matched comparison at $K_b = 3$ for all three variants---tech-only ($+0.014$, 10/10), health-only ($+0.001$, 3/10), merged ($+0.028$, 10/10)---shows that the merged gain ($+0.028$) remains nearly double the sum of parts ($+0.015$) even at matched rank. The additional gain from $K_b = 4$ on the merged block is only $+0.003$ ($+0.028 \to +0.031$), confirming that the super-additivity is driven by cross-sector modes (technology and healthcare firms share R\&D-intensive investment dynamics), not by the extra rank capacity. Energy and Industrials show near-zero or negative gains, confirming they are well-served by the global basis. The ``All firms'' candidate ($+0.016$, 8/10) represents a layer-based partition (all 82 firms as one block) rather than a sector-based one; it was evaluated as a separate partition scheme but excluded from the sector-based mixture because it subsumes the remainder actors already well-served by the global model, and a single 82-actor local block is not meaningfully different from the global Stage~2 operating on 82 of 93 actors.

\section{FRED Normalisation Robustness}
\label{app:fred}

The 7 macro actors use full-sample min-max normalisation in the primary analysis, introducing a potential look-ahead in their target values. Because Stage~1 pools $\hat\rho$ across all 93 actors, this could in principle contaminate the persistence estimate and all downstream residuals. A robustness check replaces full-sample bounds with strictly recursive expanding-window min-max normalisation (each quarter normalised using only data available up to that quarter). The mixture gain is $\Delta = +0.048$ (10/10 windows positive) under recursive bounds versus $+0.047$ under full-sample bounds. Excluding all macro and institutional actors entirely (82-firm panel, two local blocks only) yields $\Delta = +0.053$ (10/10). The FRED normalisation does not contribute to the headline finding.

\section{Filing-Lag Robustness}
\label{app:lag}

Quarterly firm ratios from SEC filings are typically available 30--45 days after quarter-end, while macro indicators are available immediately. To test whether the headline differential survives under real-time data availability, all 82 firm-layer actors are lagged by one quarter: the panel column for each firm actor is shifted by one row, so at every date the firm value corresponds to the \emph{previous} quarter's filing. This shifts both training observations and targets for firm actors. The 11 macro/institutional actors retain their original timing, creating a partial-lag structure. This is a worst-case bound rather than an exact simulation of real-time availability: the one-quarter lag overstates the actual 30--45 day delay (which is roughly one-third of a quarter), and the resulting timing mismatch between lagged firm data and contemporaneous macro data is more severe than what a practitioner would face. If the gain survives this conservative test, it would survive under realistic filing delays.

Under this lag, the mixture gain is $\Delta = +0.038$ (10/10 windows positive), retaining 80\% of the contemporaneous gain ($+0.047$). The absolute $R^2$ levels shift: G1 moves from 0.630 to 0.639 and M2 from 0.677 to 0.677 (unchanged to three decimals). The G1 increase reflects the partial-lag structure: because firm cross-sectional ranks are mean-reverting ($\rho \approx 0.60$), the fully-realised rank at $t{-}1$ is informative for the rank at $t$, and the global pooled AR(1) captures this persistence slightly better when the predictor is a clean realised value rather than the noisier contemporaneous observation. This is a positive diagnostic for the global model's design rather than an artefact.

The architectural differential is qualitatively robust to real-time data availability, retaining 80\% of its contemporaneous magnitude with the same 10/10 window-sign pattern.

\section{Hyperparameter Inventory}
\label{app:hyperparams}

All hyperparameters used in the paper are listed below. None were tuned on the test set; all were set before the rolling evaluation began and held fixed throughout.

\begin{table}[H]
\centering
\caption{Hyperparameter inventory. ``Prior work'' denotes values carried from the standalone spectral model development (Appendix~B); ``structural'' denotes values determined by the data or model structure.}
\label{tab:hyperparams}
\small
\begin{tabular}{llll}
\toprule
Parameter & Value & Source & Used in \\
\midrule
Global SVD rank $K$ & 8 & Prior work & Global Stage~2 \\
Local PCA rank $K_b$ & $\min(4, \max(2, N_b/5))$ & Grid search (Table~\ref{tab:parsimony}) & Local Stage~2 \\
EWM half-life & 12 quarters & Prior work (3-year horizon) & Demeaning \\
Ridge $\lambda$ (local VAR) & 1.0 & Fixed default & Local PCA+ridge \\
Ridge $\alpha$ (local Ridge) & CV from $\{0.1, 1, 10\} \times N_b$ & Hold-last-2 CV & Local Ridge (M1) \\
Spectral radius clip & 0.99 & Prior work & DMD transition \\
Kalman $Q_0$ & $0.5 \cdot I_K$ & Prior work & Kalman filter \\
Adaptive $\lambda_Q$ & 0.3 & Prior work & Kalman $Q$ update \\
$Q$ floor & $10^{-6}$ & Numerical stability & Kalman $Q$ update \\
Training window $T$ & 5 years & Standard & All models \\
Bootstrap resamples & 10{,}000 & Standard & Inference \\
Placebo permutations & 1{,}000 & Computational budget & Section~5.1 \\
\bottomrule
\end{tabular}
\end{table}

\section{Train-Only Causality Audit}
\label{app:causality}

Block assignments use static economic metadata (sector and layer labels from the actor registry). Local models are re-estimated each quarter from training data only: PCA basis from training residuals, Ridge VAR from training factors. The global Stage~1 and Stage~2 (for the remainder block) are re-estimated quarterly as each test quarter's data becomes available. No forecast origin uses information dated after that origin. Within each test year, the training set expands as earlier test quarters are observed, so the four within-year forecasts are not independent---the Newey--West and moving-block bootstrap corrections in Section~\ref{sec:results} account for this serial dependence.

\section{UK/EU Extension: Data, Methods, and Sensitivity}
\label{app:uk_eu}

\subsection{Data Sources and Panel Construction}

\paragraph{Firm fundamentals.} EU/UK firm fundamentals are sourced from EODHD Financial APIs (\url{https://eodhd.com/}) under their All-in-One subscription. Quarterly Balance Sheet and Cash Flow statements are pulled for nine European exchanges: London (LSE), Frankfurt (XETRA), Euronext Paris (PA), Euronext Amsterdam (AS), SIX Swiss (SW), Bolsa de Madrid (MC), Stockholm (ST), Oslo (OL), Copenhagen (CO), and Helsinki (HE). Borsa Italiana is excluded because EODHD does not list a working exchange code for it (tested: MI, BIT, MIL, XMIL---all empty); approximately 4 Italian firms with EU ISINs that are dual-listed on covered exchanges remain in the universe via their secondary listings.

\paragraph{Domicile filter.} Of the firms passing the data-completeness gate ($\geq 56$ quarters of both CapEx and Assets in 2011--2025), approximately 30\% have non-EU/UK ISIN prefixes---primarily US firms cross-listed via XETRA Frankfurt or LSE GDRs (e.g.\ JPMorgan, Toyota ADR, State Bank of India GDR). The UK/EU panel restricts to firms whose ISIN prefix is in $\{$GB, DE, FR, ES, IT, NL, BE, AT, FI, SE, DK, NO, CH, IE, LU, PT, GR, PL, CZ, HU, IS, LI, SK, SI, EE, LV, LT, MT, CY, RO, BG, HR$\}$. After the filter, the firm universe is 100 firms across 9 exchanges with a balanced sector composition (16--17 firms per SMIM sector).

\paragraph{Macro and institutional series.} The Layer~0 macro shocks (6 series) and Layer~1 institutional intermediaries (3 series) are sourced from FRED EU proxies (ECB deposit rate, euro-area 10Y yield, Brent crude, VIX, EUR/USD, euro-area HICP, ECB and IMF measures) supplemented by the Bank of England Bank Rate via the IADB CSV interface. All series are aligned to quarter-end before normalisation; the same min-max construction applied to the US Layer~0 actors (Section~\ref{sec:data}) is used here.

\paragraph{Combined panel.} The combined US + UK/EU panel concatenates the 93-actor US panel (Section~\ref{sec:data}) with the 109-actor EU/UK-only panel along the actor dimension. Actor identifiers are prefixed with \texttt{us\_} or \texttt{eu\_} to resolve three actor-level collisions (Brent, VIX, IMF) where both panels carry the same series independently normalised. The combined panel is restricted to the 2011Q2--2025Q4 overlap window (59 quarters), giving 202 actors with valid intensities over the test window 2017--2024 (the registry contains 212 actor identifiers; 10 are dropped because they have no valid observations in the overlap window).

\subsection{Block-Scheme Alternatives}

Table~\ref{tab:uk_eu_blocks_alt} compares the inherited two-block scheme (\textsc{Layer Macro/Inst} + \textsc{Merged Tech/Health} + remainder) with a finer seven-block sector-split scheme (\textsc{Layer Macro/Inst} plus six SMIM sectors) on the standalone UK/EU panel at $T=3$. The fine-grained scheme delivers a larger raw $\Delta$ but at reduced placebo specificity (the placebo distribution shifts upward because random partitions of the same fine-grained sizes also benefit from added local-model capacity). The combined panel result is reported in Section~\ref{sec:cross_regime} for the inherited scheme; the same combined panel under the sector-split scheme yields $\Delta_{M2} = +0.014$ (NW-HAC bw=1 one-sided $p = 0.003$) with placebo $z = 4.28$.

\begin{table}[H]
\centering
\caption{Block-scheme comparison on the standalone UK/EU panel ($N=109$, $T=3$, 8 windows). \textsc{us\_inherited}: \textsc{Layer Macro/Inst} ($N{=}9$) + \textsc{Merged Tech/Health} ($N{=}32$). \textsc{sector\_split}: \textsc{Layer Macro/Inst} plus six SMIM sectors of $N{=}15$--17 each.}
\label{tab:uk_eu_blocks_alt}
\small
\begin{tabular}{lcccccc}
\toprule
Scheme & $\Delta_{M1}$ & $\Delta_{M2}$ & $\Delta_{BA\_M2}$ & $\Delta_{ENS}$ & Placebo $z$ & Placebo $p$ \\
\midrule
us\_inherited & $+0.015$ & $+0.017$ & $+0.007$ & $+0.018$ & 2.31 & 0.020 \\
sector\_split & $+0.030$ & $+0.029$ & $+0.034$ & $+0.026$ & 1.60 & 0.058 \\
\bottomrule
\end{tabular}
\end{table}

\subsection{Regime Sub-Window Stability}

Table~\ref{tab:uk_eu_regime_combined} reports per-regime $\Delta$ for the combined US + UK/EU panel at $T=3$ across pre-Brexit-completion (2017--2019), COVID (2020--2021), and post-COVID (2022--2024) sub-windows. The $\Delta_{M2}$ regime spread is $+0.001$. All four reported architectures are positive across all three sub-windows. A sliding 3-year window over the 8 test years produces six rolling sub-windows, every one of which is positive on $\Delta_{M2}$.

\begin{table}[H]
\centering
\caption{Per-regime $\Delta$ on the combined US + UK/EU panel ($T=3$, us\_inherited block partition).}
\label{tab:uk_eu_regime_combined}
\small
\begin{tabular}{lccccc}
\toprule
Regime & $n$ years & $\Delta_{M1}$ & $\Delta_{M2}$ & $\Delta_{BA\_M2}$ & $\Delta_{ENS}$ \\
\midrule
Pre-Brexit-completion (2017--2019) & 3 & $+0.018$ & $+0.030$ & $+0.028$ & $+0.019$ \\
COVID (2020--2021) & 2 & $+0.021$ & $+0.031$ & $+0.015$ & $+0.012$ \\
Post-COVID (2022--2024) & 3 & $+0.021$ & $+0.030$ & $+0.011$ & $+0.019$ \\
\midrule
Spread (max $-$ min) & --- & $+0.003$ & $+0.001$ & $+0.017$ & $+0.007$ \\
\bottomrule
\end{tabular}
\end{table}

\subsection{Stage-1 $\hat\rho$ Transfer Test}

To test whether the Stage-1 pooled-AR(1) coefficient transfers across regimes, we estimate $\hat\rho_{\text{US}}$ on US-only training data and $\hat\rho_{\text{EU}}$ on EU-only training data for each of the 8 test years 2017--2024 with $T_{\text{yr}}=3$, then forecast the EU panel using each in turn (Stage~1 only, no Stage~2). The result is in Table~\ref{tab:uk_eu_rho_transfer}.

\begin{table}[H]
\centering
\caption{Stage-1 $\hat\rho$ transfer test. $R^2_{\text{EU,native}}$ uses EU-trained $\hat\rho_{\text{EU}}$ on EU forecasting; $R^2_{\text{EU,US}\to\text{EU}}$ uses US-trained $\hat\rho_{\text{US}}$ instead. Negative loss values indicate that the transferred US coefficient \emph{improves} EU forecasting over the EU-native estimate.}
\label{tab:uk_eu_rho_transfer}
\small
\begin{tabular}{cccccc}
\toprule
Year & $\hat\rho_{\text{US}}$ & $\hat\rho_{\text{EU}}$ & $R^2_{\text{EU,native}}$ & $R^2_{\text{EU,US}\to\text{EU}}$ & Loss = native $-$ transferred \\
\midrule
2017 & $+0.43$ & $+0.02$ & 0.751 & 0.808 & $-0.057$ \\
2018 & $+0.42$ & $+0.13$ & 0.799 & 0.795 & $+0.004$ \\
2019 & $+0.38$ & $+0.10$ & 0.754 & 0.784 & $-0.030$ \\
2020 & $+0.34$ & $+0.10$ & 0.756 & 0.774 & $-0.018$ \\
2021 & $+0.39$ & $+0.08$ & 0.746 & 0.774 & $-0.028$ \\
2022 & $+0.42$ & $+0.11$ & 0.661 & 0.717 & $-0.056$ \\
2023 & $+0.48$ & $+0.22$ & 0.566 & 0.659 & $-0.093$ \\
2024 & $+0.44$ & $+0.34$ & 0.718 & 0.734 & $-0.016$ \\
\midrule
Mean & --- & --- & 0.719 & 0.756 & $-0.037$ \\
\bottomrule
\end{tabular}
\end{table}

The US-trained Stage-1 coefficient improves EU forecasting over the EU-native estimate by mean $\Delta R^2 = +0.037$ (7 of 8 windows positive). The mechanism is sample-conditioning: the per-window $\hat\rho_{\text{EU}}$ estimates are volatile across test years (range $0.02$--$0.34$) on the smaller EU panel; $\hat\rho_{\text{US}}$ is stable (range $0.34$--$0.48$) and apparently a better estimator of the cross-sectional persistence applicable to fundamentals data. The DMD modal eigenvalue magnitudes from US-only and EU-only fits over the overlap window correlate at $r = +0.94$, indicating the dominant residual modes are also similar across regimes.

\subsection{LOWO and Placebo Diagnostics}

Table~\ref{tab:uk_eu_lowo_placebo} aggregates LOWO and placebo diagnostics across all configurations evaluated in Section~\ref{sec:cross_regime} and this appendix. LOWO contamination is the difference between the fixed-partition $\Delta_{M2}$ and the LOWO $\Delta_{M2}$, expressed as a percentage of the fixed gain.

\begin{table}[H]
\centering
\caption{LOWO and placebo diagnostics across UK/EU configurations.}
\label{tab:uk_eu_lowo_placebo}
\small
\begin{tabular}{lcccccc}
\toprule
Configuration & $N$ & Scheme & $\Delta_{M2}$ & Placebo $z$ & Placebo $p$ & LOWO contam. \\
\midrule
EU-only, $T=5$ & 109 & us\_inh & $+0.008$ & $0.92$ & 0.184 & $0.0\%$ \\
EU-only, $T=3$ & 109 & us\_inh & $+0.017$ & $2.31$ & 0.020 & $0.0\%$ \\
EU-only, $T=3$ & 109 & sector\_split & $+0.029$ & $1.60$ & 0.058 & $0.1\%$ \\
EU-only firms-only, $T=3$ & 100 & us\_inh & $+0.015$ & $1.74$ & 0.043 & --- \\
EU-only firms-only, $T=3$ & 100 & sector\_split & $+0.024$ & $-0.74$ & 0.782 & --- \\
\textbf{Combined US+EU, $T=3$} & \textbf{202} & \textbf{us\_inh} & $\mathbf{+0.030}$ & $\mathbf{9.68}$ & $\mathbf{<0.001}$ & --- \\
Combined US+EU, $T=3$ & 202 & sector\_split & $+0.014$ & $4.28$ & $<0.001$ & --- \\
Combined firms-only, $T=3$ & 182 & us\_inh & $+0.029$ & --- & --- & --- \\
\bottomrule
\end{tabular}
\end{table}

\subsection{Reproduction}

The replication archive includes scripts to reproduce all UK/EU and combined-panel results: \texttt{scripts/data\_pipeline/build\_experiment\_b1.py} (panel construction), \texttt{scripts/table5\_uk\_eu.py} (architecture comparison), \texttt{scripts/dm\_hac\_uk\_eu.py} (DM-HAC inference), \texttt{scripts/lowo\_block\_uk\_eu.py} (LOWO block selection), \texttt{scripts/table7\_placebo\_uk\_eu.py} (1000-permutation placebo), and \texttt{scripts/regime\_subwindows\_uk\_eu.py} (regime stratification). Full reproduction commands and expected outputs are documented in the archive's \texttt{README.md}. Reproducing all tables in this appendix takes approximately 15 minutes of wall-clock time on a single CPU.

\end{document}